\documentclass[twocolumnprb,amsmath,amssymb,showpacs,lengthcheck]{revtex4-1} %revtex4-1

\usepackage{graphicx}
\newcommand{\bvec}{\mathbf}
\begin{document}

\title{Impact ionization dynamics in small band-gap 2D materials from a coherent phonon mechanism}
\author{Stephan Michael}
\author{Hans Christian Schneider}
\affiliation{Department of Physics and Research Center OPTIMAS, TU Kaiserslautern, P.O. Box 3049, 67653 Kaiserslautern, Germany}

\begin{abstract}
We study theoretically the role of carrier multiplication due to impact ionization after an ultrafast optical excitation in a model system of a quasi-two dimensional material with a small band gap. As a mechanism for the photo-induced band gap narrowing we use coherent phonons, which mimics the quenching of an insulator phase. We discuss the importance of impact ionization in the ultrafast response, and investigate the interplay between carrier and band dynamics. Our model allows us to compare with recent experiments and identify signatures of carrier multiplication in typical electronic distribution curves as measured by time-resolved photoemission spectroscopy. In particular we investigate the influence of the shape of the bands on the carrier multiplication and the respective contributions of band and carrier dynamics to electronic distribution curves.
\end{abstract}

\maketitle

\section{Introduction}
\label{Introduction}

Recent developments in time- and angle-resolved photoemission spectroscopy (trARPES) have opened up the possibility to study the material response after ultrafast optical excitation using photoemission techniques.~\cite{Rohwer2011,Sobota2014,Eich2014} This progress has facilitated the study of correlated and nanoscale quantum materials.~\cite{Smallwood2016} Besides graphene,~\cite{Gierz2016,Gierz2013,Stroucken2013,Mihnev2016,Kadi2015} other two-dimensional materials,~\cite{Bhimanapati2015} in particular transition-metal dichalcogenides (TMDC),~\cite{Chhowalla2013,Voiry2015,Steinhoff2016} have been the center of current investigations. Associated with this topic is an active interest in metal-insulator transitions.~\cite{Hellmann2012} Besides Mott insulators~\cite{ghatak2011nature,cheiwchanchamnangij2012quasiparticle,chernikov2015population,chen2016ultrafast,Steinhoff2016,steinhoff2017excitons,moon2018soft} these transitions can appear as excitonic and Peierls insulators, where a formation of a charge density wave (CDW) and a periodic lattice distortion occur.~\cite{Gruner1988} However, charge density waves are observed in many solids and their origin is still under debate.~\cite{Zhu2015,Flicker2015,Cho2016,leroux2018traces,nakata2018anisotropic} Further, changing the symmetry of a material via optically induced phase transitions offers new ways to manipulate material properties on ultrafast timescales.~\cite{Wall2012,mizokawa2009local} Research about the ultrafast response of two dimensional materials like TMDC in connection with their rich electronic phase diagrams, e.g., superconductivity~\cite{Sykora2009,Kim2015} or CDW phases, may well be important for understanding the basic physics for the design of future ultrafast (optoelectronic or optospintronic) devices.~\cite{Ritschel2015,Gu2016,khitun2016two} 

Materials like 1T-TaS$_{2}$, 2H-TaSe$_{2}$, and 1T-TiSe$_{2}$ have been studied in some detail,~\cite{Rossnagel2011,Porer2014,zhang2016unveiling} but there still is a controversy about the origin of the CDW phases in those materials, especially for 1T-TiSe$_{2}$. In Refs.~\onlinecite{Monney2009,Monney2010,Cazzaniga2012,Monney2012,Li2007,Monney2011,kogar2017signatures,chen2017reproduction,pasquier2018excitonic,golez2016photoinduced} an excitonic insulator mechanism was identified. However, there are also arguments that an electron-lattice interaction with the help of the Jahn-Teller effect leads to a Peierls-like CDW transition and the accompanying opening of a gap.~\cite{Rossnagel2002,karam2018strongly,hughes1977structural,hellgren2017critical} To our knowledge, the prevailing explanation is a combination of exciton-formation and electron-phonon coupling.~\cite{Kidd2002,VanWezel2010,Monney2012a,VanWezel2010a,Phan2013,Watanabe2015,Bianco2015,CMonney2016,kogar2017signatures,chen2017reproduction,pasquier2018excitonic,karam2018strongly,maschek2016superconductivity,singh2017stable,hellgren2017critical} In this context, the chirality of the CDW~\cite{Ishioka2010,Castellan2013,silva2018elemental,hildebrand2018local} and a softening of  phonon modes~\cite{Weber2011,maschek2016superconductivity} have also been discussed. 

The present paper is devoted to a study of non-equilibrium carrier dynamics during an optically induced phase change between an ``insulator'' and a ``metallic'' phase in a system with a small band gap, where carrier-scattering processes may lead to carrier multiplication due to impact ionization. On the basis of experimental results and a simple model calculation it has recently been argued in Ref.~\onlinecite{Mathias2016} that in 1T-TiSe$_{2}$ excitation by an ultrafast optical pulse induces carrier multiplication and gap-closing dynamics, which amplify each other during the quenching of the CDW phase. In this paper, we investigate theoretically in more detail this interplay between carrier dynamics (in particular, carrier multiplication) and quasi-particle band-structure change, i.e., gap quenching. We employ a dynamical model that is capable of describing aspects of the ultrafast response of small band-gap 2D materials. We assume an electron-phonon based mechanism behind the formation of the CDW state, but we do not attempt a microscopic description of the complete change between insulator and metallic phase. Instead, we restrict our attention to the onset of the phase transition starting from the CDW insulator phase, and model the relevant lattice dynamics by coherent phonons. These coherent phonons interact with the optically excited electronic dynamics and, in turn, change the quasi-particle band structure via a modulation of hybridization between electronic orbitals centered at the ions that oscillate with the coherent phonon. An important goal of this paper is to study the interplay between carrier multiplication effects and  quasi-particle band-structure dynamics by dynamical calculations for a concrete mechanism. In particular we explore the consequences for quantities accessible in recent experiments, where carrier multiplication and band-strucure dynamics cannot easily be disentangled.~\cite{Mathias2016} In particular, we study the influence of different excitation scenarios, and compare the results for different band structures (parabolic and Mexican-hat shaped bands).

The outline of this paper is as follows. In Sec.~\ref{sec:QP-bands} we first introduce a model composed of a tight-binding band structure and carrier-phonon interaction in which the quenching of the insulator phase is due to the coupling to coherent phonons. In a nonequilibrium situation, this electron-phonon coupling results in the change in quasi-particle bands associated with a Peierls-like transition. In Sec.~\ref{sec:carrier-dynamics} we set up the equations of motion for the relevant distribution functions including the optical-excitation contribution and Coulomb interaction. Numerical results are presented in Sec.~\ref{sec:numerical-results}. We discuss here in particular the influence of model parameters on the carrier dynamics the interplay of carrier multiplication and gap-closing dynamics and their signatures in electronic distribution curves. Technical details concerning the tight-binding model and the numerical solution of the dynamical equations including the gap dynamics are collected in Appendices~\ref{app_tb} and~\ref{app_eom}. We conclude the paper in Sec.~\ref{sec:conclusion}.

\section{Quasi-particle Electronic Structure Calculation}
\label{sec:QP-bands}

As we want to describe carrier dynamics that accompany the quenching of a small band-gap insulator phase, we first need to address how the quasiparticle band structure changes during this phase transition. While a variety of different models for charge-density-wave insulators exist,~\cite{Zhu2015} the classification for materials like 1\emph{T}-TiSe$_2$ or 1T-TaS$_2$ is not straightforward. Among other reasons, this is because electron-electron and electron-phonon interactions may both play an important role in the phase transition dynamics. As we focus here on the carrier dynamics, it is beyond the scope of this paper to include such complex interdependencies. Instead of determining the insulator phase from the normal phase, we start from a TB model of the band structure in the insulator phase and describe the quenching of this phase as an effective misalignment of the atomic positions of the different atoms in the unit cell. This effective atomic displacement after an optical excitation enters our calculation as a coherent phonon. 

\subsection{Tight-Binding Model}
\label{sec:tightbinding}

We employ a tight-binding model to describe a quasi-two dimensional material with two atomic species A$_{d}$ and A$_{p}$ in the insulator phase. The parameters are chosen such as to reproduce some important characteristics of electronic states in TMDCs. In the case of a TMDC, the atomic species A$_{d}$ is the transition metal (e.g. Ti) with d-type or f-type valence orbitals and the atomic species A$_{p}$ is the chalcogen (e.g., Se) with p-type valence orbitals. For instance, in Refs.~\onlinecite{VanWezel2010,VanWezel2010a} it was found that for 1T-TiSe$_{2}$ only the three hopping parameters dd$_{\sigma}$, pp$_{\sigma}$ and pd$_{\pi}$ contribute significantly to the behavior of states with energies close to the Fermi energy. This allows one to use a restricted model that includes only these hopping parameters.

As we do not attempt a microscopic model of the physics underlying the phase transition and as we are mainly interested in the electronic dynamics close to the small band gap, which is the indicator of the CDW and typically opens at high symmetry points, such as $\Gamma$ or $M$ points, we  use a simple two-band tight-binding model to capture the characteristics of the carrier and band dynamics around the gap after an ultrashort optical excitation. We explain the relation of this ansatz to existing tight-binding models of transition-metal dichalcogenides in Appendix~\ref{app_tb}. For now, we take the model tight-binding hamiltonian in the form  
\begin{equation}
	\begin{split}
		H_{\text{TB}} &= \epsilon^{p}_{0} \big(c^{ps}_{\bvec{k}}\big)^{\dag} c^{ps}_{\bvec{k}} + \epsilon^{c}_{0} \big(c^{ds}_{\bvec{k}}\big)^{\dag} c^{ds}_{\bvec{k}} \\ 
		& + 2 V_{\text{pp}} \big[ \cos(k_{x} e_{x}) + \cos(k_{y} e_{y}) \big]
			(c^{ps}_{\bvec{k}})^{\dag} c^{ps}_{\bvec{k}} \\ 
		& + V_{\text{pd}}  (c^{ps}_{\bvec{k}})^{\dag} c^{ds}_{\bvec{k}} 
			+ V_{\text{dp}}  c^{ds \dag}_{\bvec{k}}bc^{ps}_{\bvec{k}} \\ 
		& + 2 V_{\text{dd}} \big[ \cos(k_{x} e_{x}) + \cos(k_{y} e_{y}) \big] (c^{ds}_{\bvec{k}})^{\dag} c^{ds}_{\bvec{k}} 
	\end{split}
\label{tb1}
\end{equation}
where $\epsilon^{p}_{0}$, $\epsilon^{d}_{0}$ are the on-site energies, and $V_{\text{pp}}$, $V_{\text{pd}}$, $V_{\text{dp}}$, $V_{\text{dd}}$ are the tight-binding coupling-elements and $\bvec{e}$ is the distance vector between two neighboring unit cells. As we do not include spin-orbit coupling, we do not explicitly write out the spin-dependence s in the following.

This model yields a conduction band mainly consisting of a d-type transition metal orbital and a valence band mainly originating from a p-type chalcogen orbital. In the neighborhood of this point the band structure has the shape of a Mexican hat with a small band gap and pronounced band mixing for the  model parameters chosen here, see Sec.~\ref{sec:numerical-results}. Close to the high symmetry point the band structure possesses rotational symmetry. We stress that the simplicity of this model and the high symmetry are not too restrictive,  because a fast angular redistribution of carriers due to electron-phonon scattering~\cite{Mittendorff2014} will smooth out the effects of anisotropy, and our results should also be transferable to non-parabolic band structures.

\subsection{Quasi-particle band dynamics and the effective Hamiltonian}
\label{sec:coherent-phonons}
%---------------------------------------------------------------------
This subsection is concerned with determination of the carrier states that accompany the onset of the phase change and that we will sometimes refer to simply as ``band dynamics''. We do not attempt a microscopic ab-initio description of the coupled electron-ion system and the transition from normal phase transition to charge-density wave phase. Instead, we use an effective hamiltonian for the system in the charge-density wave state that already incorporates the lattice distortion induced by the electron-phonon interaction. In this model, the coherent phonon leads to an ionic displacement that causes a change of the hybridization between electronic orbitals centered at different ions, which changes the effective hamiltonian. Due to the dependence on the phonon dynamics, the effective hamiltonian becomes time dependent. 

We begin with the free phonon Hamiltonian for the coherent phonon, i.e. $\bvec{q}=\bvec{0}$,
\begin{equation}
H_{\text{Ph}} = \hbar \omega_{0} \big( b_{\bvec{0}} b^{\dag}_{\bvec{0}} + \frac{1}{2} \big)
\label{H_ep0}
\end{equation}
describes a coherent phonon (cpn) with $\bvec{q}=\bvec{0}$ that leads to a distortion consistent with the symmetry of the material, e.g., the A1g mode, in which the two kinds of atoms are displaced in the unit cell. %% Cite Wezel: Dreier-Schwingung 
The coherent phonon couples to the electrons by modulating the p-d hybridization 
\begin{equation}
	H_{\text{e-cpn}} = \sum_{\bvec{k}} g^{pd}_{\bvec{0}} 
		( b_{0} + b^{\dag}_{0}) (c^{p}_{\bvec{k}+\bvec{0}})^{\dag}c^{d}_{\bvec{k}} 
		+ \text{h. c.} 
\label{H_ep1}
\end{equation}
where $g^{pd}_{\bvec{0}}$ is the matrix element for the coupling of electrons to the coherent phonon in the orbital basis. We assume for simplicity that this matrix element is $\bvec{k}$-independent. 

The interaction of electrons with a coherent phonon includes a mean-field contribution  
%SM: SM Version
\begin{equation}
	H_{\text{e-cpn}}^{\text{(mf)}} = \sum_{\bvec{k}} \sum_{l_{1},l_{2}} g^{l_{1}l_{2}}_{0} (B_{0} + B^{\dag}_{0}) c^{l_{1} \dag}_{\bvec{k}+\bvec{0}}c^{l_{2} }_{\bvec{k}} + \text{h.c.} 
\label{H_ep2}
\end{equation}
where $B_{0}=\langle b_{0} \rangle$ is the coherent phonon amplitude, and $l_{1},l_{2} \in \{ p,d \}$ denotes the orbital index.
The mean-field part of the coupling hamiltonian to the coherent phonon does not contain phonon operators and can be combined with $H_{\text{TB}}$ to an effective hamiltonian for the carrier system that describes the states around the Fermi energy
\begin{equation}
H_{\text{eff}}= H_{\text{TB}} + H^{\text{(mf)}}_{\text{e-cpn}}
\label{H_ep4}
\end{equation}
As $H_{\text{eff}}$ is time dependent its eigenvalues $\epsilon_{b,\bvec{k}}$ and eigenvectors $\Psi_{b,\bvec{k}}(\bvec{ r})$ are calculated for every time-step of the dynamical calculation. Thus, matrix elements $g^{b_{1} b_{2}}_{0}$ and $\rho^{b_{1}b_{2}}_{\bvec{k}}$ generally involve time-dependent basis states as will be discussed in Appendix~\ref{app_eom}. In this time-dependent eigenbasis, $n^{b}_{\bvec{k}} = \rho^{bb}_{\bvec{k}}$ can be interpreted as the occupation of the state $|b,\bvec{k} \rangle$ at that time and $g^{b_{1} b_{2}}_{0,\bvec{k}}$ the corresponding phonon matrix element. In particular, the matrix element $g^{pd}_{0}$ in the orbital basis is related to matrix elements $g^{cc}_{0,\bvec{k}}$ and $g^{\mathrm{v}\mathrm{v}}_{0,\bvec{k}}$ in the time-dependent basis. Assuming that the coherences in this equation-of-motion die out faster than the dynamics of interest, we obtain the equation of motion 
\begin{equation}
\begin{split}
	\frac{d}{dt} B_{\bvec{0}} =& - ( i \omega_{\bvec{0}} + \gamma^{P}_{\text{deph}}) B_{\bvec{0}} +
	\frac{1}{i\hbar} \sum_{\bvec{k}} \sum_{b} 
		\big[ (g^{bb}_{0,\bvec{k}})^* n^{b}_{\bvec{k}} \big] \\
	 =  & - ( i \omega_{\bvec{0}} + \gamma^{P}_{\text{deph}}) B_{\bvec{0}}\\ 
		& + \frac{1}{i\hbar} \Big(  \sum_{\bvec{k}} (g^{cc}_{0,\bvec{k}})^* n^{c}_{\bvec{k}} 
		   + \sum_{\bvec{k}} (g^{vv}_{0,\bvec{k}})^* n^{v}_{\bvec{k}} \Big) 
\end{split}
\label{H_ep5a}
\end{equation} 
The coupling matrix elements $g^{pd}_{0}$, which are off diagonal with respect to the orbital index, influence the band occupations $n^{b}_{\bvec{k}}$ via $g^{bb}_{0,\bvec{k}}$ matrix elements, which are diagonal with respect to the band index, and thus drive the coherent phonon amplitude Eq.~\eqref{H_ep5a}.

\section{Carrier dynamics via equation of motion technique}\label{sec:carrier-dynamics}
%=======================================================================================
\subsection{Optical excitation}\label{sec:optical-excitation}
%-------------------------------------------------------------
We model the optical excitation after a recent experiment on 1T-TiSe$_{2}$ in Ref.~\onlinecite{Mathias2016}, where carriers were excited with an 1.6 eV pulse around 200 meV above the Fermi level into a Ti 3d band around the M-point. Around this high symmetry point, only a small band-gap exists between the Ti 3d band and a back-folded Se 4p band. As the holes, which are likely created in a Se 4p(x,y) bands, never appear close to the Fermi surface, we do not include these band states in our two-band tight-binding model. Further, the dispersions of the bands of interest are different (i.e., have very different curvature in our simplified case), so that in the first few hundred femtoseconds the excited holes have no chance to reach the Fermi surface and no efficient contribution to the ultrafast carrier and band response around Fermi surface is possible, as found in experiment.~\cite{Mathias2016}
Thus, we model the optical excitation between the conduction band ``c'' mainly originating from the d-type orbital of atom species A$_{d}$ and a third band $\mathrm{v}^{\prime}$ below the Fermi surface by
\begin{equation}
\begin{split}
	\frac{d}{dt} p^{b_{1} b_{2}}_{\bvec{k}} \Big|_{\text{opt}} =& - \left(i \omega^{b_{1} b_{2}}_{\bvec{k}} + \gamma^{P}_{\text{deph}}\right) p^{b_{1} b_{2}}_{\bvec{k}}\\
	&- i \Omega^{b_{1} b_{2}}_{\bvec{k}}\left( n^{b_{1}}_{\bvec{k}} - n^{b_{2}}_{\bvec{k}} \right)
\end{split}
\end{equation}
and
\begin{equation}
	\frac{d}{dt} n^{b_{1}}_{\bvec{k}}\Big|_{\text{opt}} =- \left( i \Omega^{b_{1} b_{2}}_{\bvec{k}} p^{b_{1} b_{2}}_{\bvec{k}} + \text{h.c.} \right)
\label{optex1}
\end{equation}
where $b \in \{ \mathrm{c},\mathrm{v}' \}$, and we have again suppressed the spin index. 

The major contribution to the coherent phonon amplitude dynamics originates from the excitation of electrons into the conduction band. This is in accordance to situations, where optical excitation can trigger a displacive A1g CDW amplitude mode by exciting electrons from bonding to antibonding states, e.g., in 1\emph{T}-TiSe$_{2}$.~\cite{CMonney2016} It is also supported by other investigations, which have found that the A1g mode shows a strong coupling to conduction electrons.~\cite{Donkov2009,Chakraborty2012} While the effects of excitonic contributions likely have to be included to obtain quantitative agreement (e.g., for the speed of the gap dynamics),~\cite{Mathias2016} the qualitative picture of the onset of a phase transition due to ultrafast optical excitation can be described by coherent phonons. In such a model, carrier multiplication also has an contribution to the dynamics of the coherent phonon amplitude, as we show in the following. 

\subsection{Carrier-carrier Coulomb scattering}
\label{sec:Coulomb-interaction}

The Coulomb scattering to describe the carrier dynamics in the first few hundred femtoseconds after the ultrafast optical excitation is also included in the equation of motion for the density matrix. As the Coulomb interaction leads to transitions between quasiparticle states, which change dynamically, we use time-dependent Bloch states. This entails not only the correction of the band-energies but also a re-calculation of the interaction-matrix elements. In general, it is associated with a transformation of diagonal density-contributions $n^{b}_{\bvec{k}}$ into off-diagonal coherence-contributions in conjunction with correlated correction-terms in the equation-of-motion. This general consequences are described and the level of approximation for the system under investigation is explained in appendix~\ref{app_eom}, where we assume a sufficiently high dephasing for these coherences, which is likely for the system under investigation, and hence the off-diagonal coherence-contributions in conjunction with correlated correction-terms in the equation-of-motion can be neglected. Thus, we implement the time-dependent basis in the description of the carrier dynamics using time-dependent band-energies and wave-functions including time-dependent Coulomb-matrix elements due to the basis transformation with of-course a time-dependent screening. Importantly, the band dynamics here leads to an additional re-distribution of carriers into the new equilibrium distribution and changes the ratio between intra and interband scattering pathways.  

The derivation of the Coulomb scattering equations itself can be established in various ways. For instance, with cluster expansion techniques or with the Green's function technique under the use of the Kadanoff-Baym equations, by applying the second-order Born approximation for the self-energy.\cite{Steinhoff2016} For the carrier-carrier scattering we neglect coherences and obtain the following equation of motion for the Coulomb scattering in Markov approximation  
\begin{equation}
\frac{d}{dt} n^{b}_{\bvec{k}} = \frac{2\pi}{\hbar}
\sum_{\bvec{k}_{2}\bvec{k}_{3}}
\sum_{b_{2}b_{3}b_{4}}  \widehat{W} \left( N^{\text{in}} - N^{\text{out}} \right) \delta \left( \Delta\epsilon \right) 
\label{carrierdyn1}
\end{equation}
with 
\begin{align}
\widehat{W} =& W^{b_{}b_{2}b_{3}b_{4}}_{\bvec{k}_{}\bvec{k}_{2}\bvec{k}_{3}\bvec{k}_{4} } \left( W^{b_{}b_{2}b_{3}b_{4}\text{*}}_{\bvec{k}_{}\bvec{k}_{2}\bvec{k}_{3}\bvec{k}_{4} } -
	W^{bb_{2}b_{4}b_{3}\text{*}}_{\bvec{k}_{}\bvec{k}_{2}\bvec{k}_{4}\bvec{k}_{3}} \right) \\
	N^{\text{in}} =& \left( 1 - n^{b}_{\bvec{k}} \right) n^{b_{2}}_{\bvec{k}_{2}}
	\left( 1 - n^{b_{3}}_{\bvec{k}_{3}} \right) n^{b_{4}}_{\bvec{k}_{4}} \\
	N^{\text{out}} =& n^{b}_{\bvec{k}} \left( 1 - n^{b_{2}}_{\bvec{k}_{2}} \right)
	n^{b_{3}}_{\bvec{k}_{3}} \left( 1 - n^{b_{4}}_{\bvec{k}_{4}} \right) \\
	\Delta\epsilon =& \epsilon^{b}_{\bvec{k}} -
	\epsilon^{b_{2}}_{\bvec{k}_{2}} + \epsilon^{b_{3}}_{\bvec{k}_{3}} - \epsilon^{b_{4}}_{\bvec{k}_{4}}
\end{align}
where $W^{b_{} b_{2} b_{3} b_{4}}_{\bvec{k}_{}\bvec{k}_{2}\bvec{k}_{3}\bvec{k}_{4}}$ are the screened Coulomb-Matrix elements, $n^{b}_{\bvec{k}}$ is the carrier distribution and $\epsilon^{b}_{\bvec{k}}$ the corresponding energy on $\bvec{k}$ for the band $b \in  \{ \mathrm{c} ,\mathrm{v}\}$. The spin-index $s$ is neglected. The screened Coulomb potential is 
\begin{equation}
	W^{b_{}b_{2}b_{3}b_{4}}_{\bvec{k}_{}\bvec{k}_{2}\bvec{k}_{3}\bvec{k}_{4}}
	= \sum_{\bvec{q}} w \left( \bvec{q} \right) I^{b_{}b_{4}}_{\bvec{k}_{}\bvec{k}_{4}} \left( \bvec{q} \right) I^{b_{2}b_{3}}_{\bvec{k}_{2}\bvec{k}_{3}} \left( - \bvec{q} \right) 
\end{equation}
with the overlap integrals
\begin{equation}
	I^{b_{}b_{4}}_{\bvec{k}\bvec{k}_{4}} (	\bvec{q} ) = \int  \Psi^{b}_{\bvec{k}}(\bvec{r})^* e^{i \bvec{q} \bvec{r}} 
	\Psi^{b_{4}}_{\bvec{k}_{4}}(\bvec{r})\,d^3r
\end{equation}
and $w(\bvec{q}) = \varepsilon^{-1} (\bvec{q}) v(\bvec{q})$. Further, $\Psi^{b}_{\bvec{k}}(\bvec{r})$ are the eigenfunctions of the time-dependent tight-binding Hamiltonian in Eq.~\eqref{H_ep4}, $v (\bvec{q})$ is the unscreened Coulomb potential including a background dielectric constant $\varepsilon_{b}$ and $A$ the normalization area. 

In the derivation of the Coulomb scattering equations, the screened Coulomb potential can be naturally included. The time-dependent screening of the Coulomb interaction is taken into account using the static limit of Lindhard dielectric function 
\begin{equation}
	\varepsilon(\bvec{q}) =  1 - 
	\frac{1}{A} \sum_{\bvec{k},b} V^{bbbb}_{\bvec{k},\bvec{k}-\bvec{q},\bvec{k},\bvec{k}-\bvec{q}} \frac{n^{\lambda}_{\bvec{k}-\bvec{q}}-n^{\lambda}_{\bvec{k}}}{\epsilon^{\lambda}_{\bvec{k}-\bvec{q}} -\epsilon^{\lambda}_{\bvec{k}}} 
\label{carrierdyn2}
\end{equation}
where $V^{b_{}b_{2}b_{3}b}_{\bvec{k}_{}\text{,}\bvec{k}_{2}\text{,}\bvec{k}_{3}\text{,}\bvec{k}_{4}}$ are the Coulomb-matrix elements calculated from the unscreened Coulomb potential $v \left( \bvec{q} \right)$.

\section{Numerical results}\label{sec:numerical-results}
%--------------------------------------------------------
\begin{figure}[tb]
\centering
\includegraphics[scale=0.6]{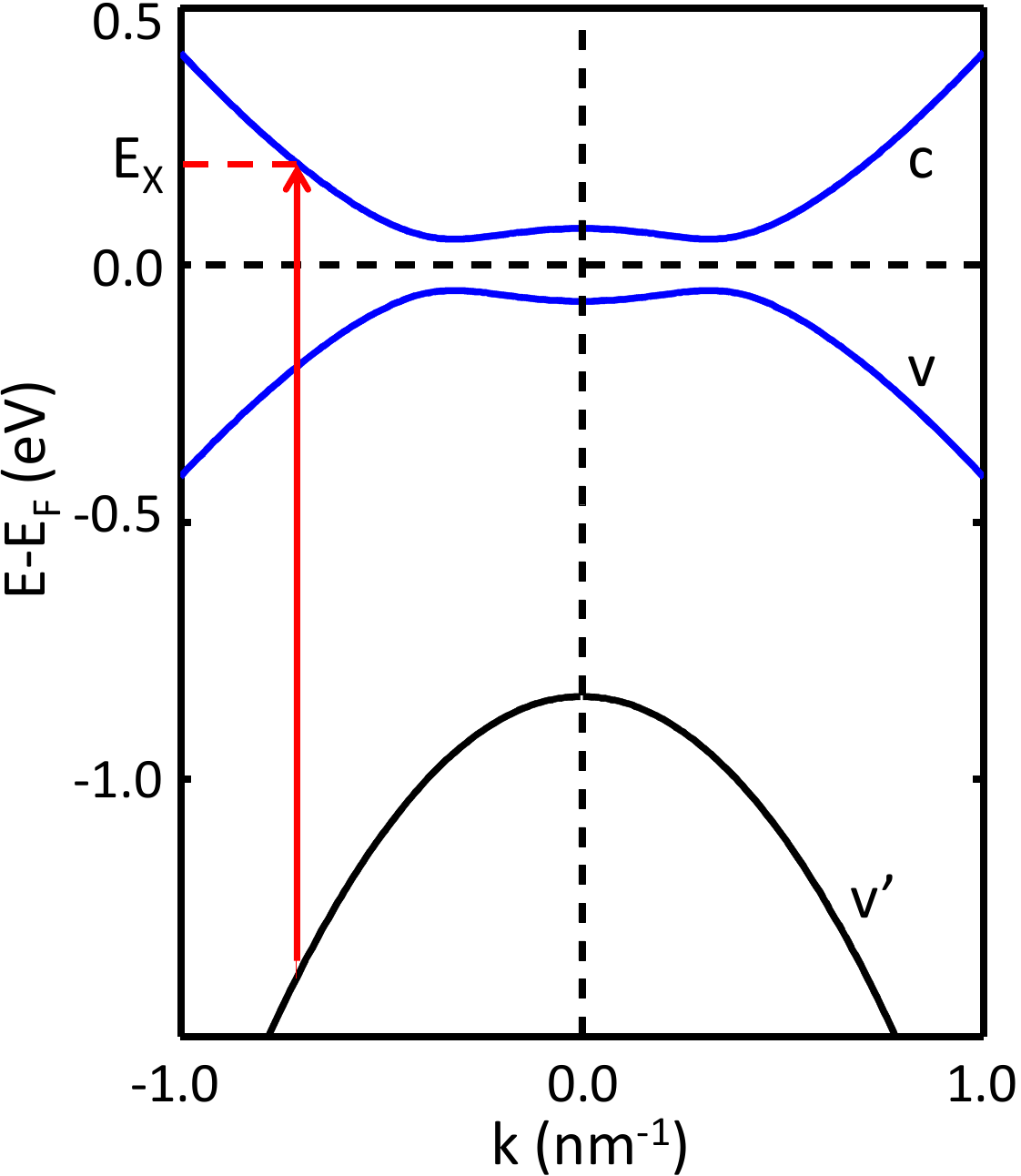}
\caption{Band structure of the unexcited material on a high symmetry point in the Brillouin zone on the Fermi surface in radial in-plane direction. The blue curves are the conduction band $c$ and the valence band $v$ calculated from the tight-binding Hamiltonian. The black curve is a third band $v\prime$ not included in the two-band tight-binding model far bellow the Fermi surface used for the ultrafast optical excitation with an 1.6 eV pulse (red arrow). The vertically dashed line is the Fermi surface and the horizontal dashed line is the high symmetry point in the Brillouin zone.}
\label{figure0}
\end{figure}

To investigate the ultrafast response of a small band-gap 2D material after an ultrafast optical excitation, and particularly of the role of impact ionization and carrier multiplication in the conduction band, we assume the setup shown in Fig.~\ref{figure0}, which we discuss here first.
From the tight-binding Hamiltonian Eq.~\eqref{tb1} from Sec.~\ref{sec:tightbinding}, we obtain two Mexican-hat shaped bands close to the Fermi surface, a conduction band ``c'' and a valence band ``v''. Further, we assume a lattice temperature of 100 K and the band gap is measured as the nearest distance between the two Mexican-hat shaped bands, which is 100 meV for the unexcited band structure. The lattice temperature is below the transition temperature and the band-gap is adapted to that of the CDW phase of a typical material like TiSe$_{2}$ as reported for example in Ref.~\onlinecite{Monney2010}. 
The optical excitation is modeled as excitation originating from a third band ``$\mathrm{v}'$'' not included in the two-band tight-binding model as explained in Sect.~\ref{sec:optical-excitation}. As indicated in Fig.~\ref{figure0}, the $\mathrm{v}'$ band is far below the Fermi surface and the ultrafast optical excitation by a pulse with a 1.6 eV photon energy excites carriers into the conduction band around 200 meV above the Fermi surface as measured by trARPES experiments on TiSe$_{2}$ reported in Ref.~\onlinecite{Mathias2016}.
For the unexcited material we assume a Fermi distribution and thus we obtain a nearly empty conduction band with negligible band corrections due to coherent phonons, see Sect.~\ref{sec:coherent-phonons} and weak screening. After the ultrashort optical excitation, effects such as band dynamics induced by coherent phonons and a carrier redistribution due to Coulomb scattering, see Sect.~\ref{sec:Coulomb-interaction},  will determine the material response as investigated in the following.   

\subsection{Characteristics of band- and carrier response}

\begin{figure}[tb]
\centering
\includegraphics[scale=0.6]{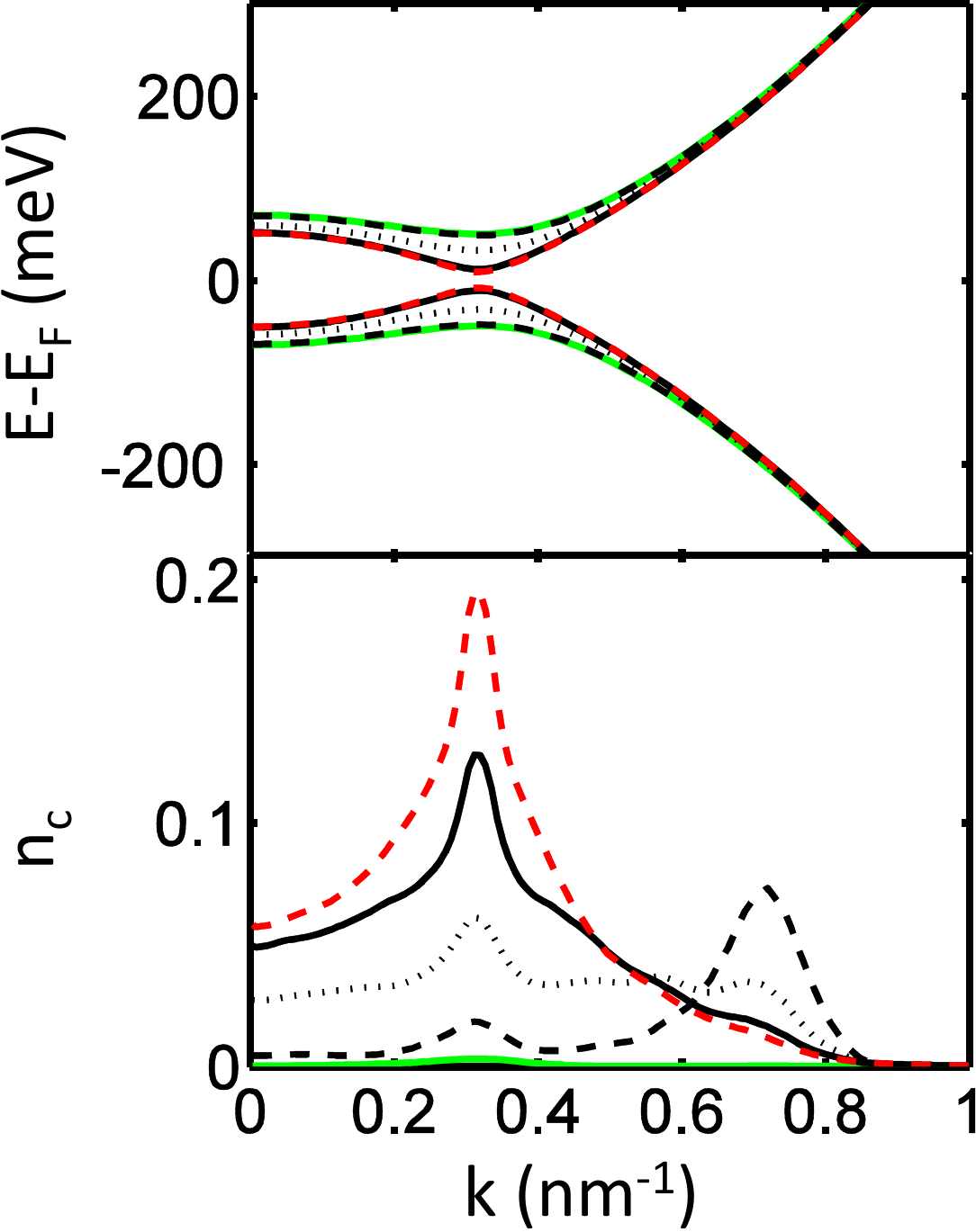}
\caption{Band dynamics (top) and carrier distribution in the conduction band (bottom) at $-25$\,fs (solid green), $25$\,fs (dashed black), 100 fs (dotted black), 175 fs (solid black), 250 fs (dashed red) after an ultrashort optical excitation by a $\sigma_T =14$\,fs pulse centered at 0 fs. Between -25 fs and 25 fs the optical excitation mainly determines the carrier dynamics and the non-equilibrium carrier distribution lead to a closing of the gap. After 25 fs the optical does not contribute any more. Now the carrier scattering and the effect of gap-closing increase the impact ionization, which contributes to a further gap-closing until a quasi-equilibrium after about 200 fs is reached.}
\label{figure1}
\end{figure}

\begin{figure}[tb]
\centering
\includegraphics[scale=0.6]{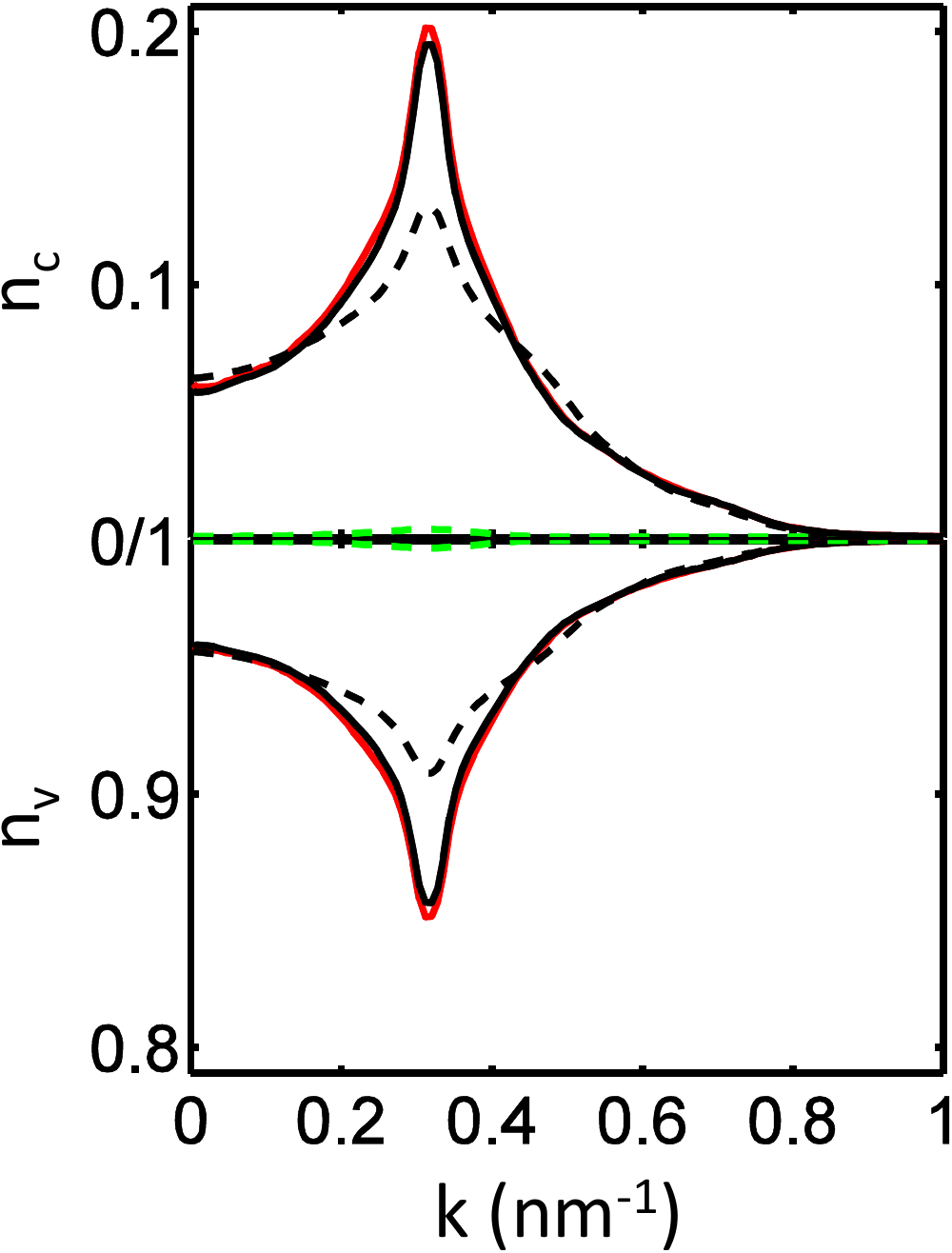}
\caption{Carrier distribution in the conduction and valence band before and long after the pulse.  $t=-25$\, fs (green dashed line) and $t= 250$\,fs for the setup with $g^{pd}_{0}=10.0$~meV (solid black line), $g^{pd}_{0}=8.0$~meV (dashed black line) and $\tilde g^{\mathrm{cc}}_{0}=8.0$\,meV (solid red line). As electrons are optically excited into the conduction band, the valence band dynamics are exclusively due to interband scattering, i.e., carrier multiplication. The setups with $g^{pd}_{0}=10.0$~meV and $\tilde g^{\mathrm{cc}}_{0}=8.0$\,meV exhibit similar carrier distributions after 250 fs.}
\label{figure3}
\end{figure}

\begin{figure}[tb]
\centering
\includegraphics[scale=0.6]{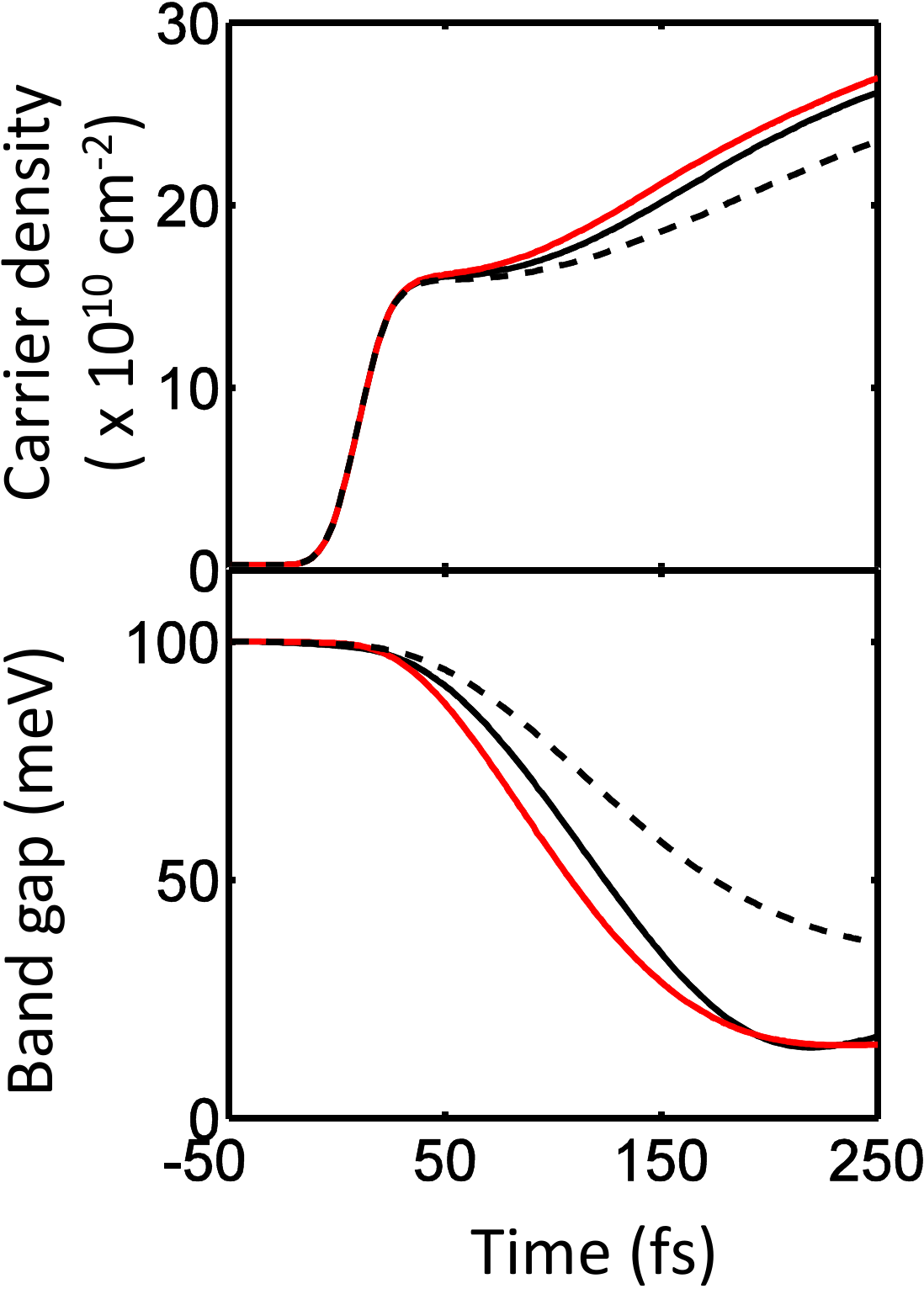}
\caption{Gap closing (bottom) and carrier density in the conduction band (top) vs. time. After the optical excitation an induced gap closing and a delayed carrier multiplication correlated to the gap closing is visible for the setup with $g^{pd}_{0}=8.0$~meV (dashed black line), $g^{pd}_{0}=10.0$~meV (solid black line), and $\tilde g^{\mathrm{cc}}_{0}=8.0$\,meV (solid red line). The setup with $g^{pd}_{0}=10.0$~meV and $\tilde g^{\mathrm{cc}}_{0}=8.0$\,meV have a similar time evolution of the carrier density and the gap closing. Between 100 fs and 200 fs (short times after optical excitation) only small derivations are visible, which vanish after 200 fs.}
\label{figure4}
\end{figure}

First, we discuss the essential characteristics of the dynamical results for the excitation described above. A technical aspect is to clarify and investigate the effect of the basis transformation of the electron-phonon matrix element between the time-dependent basis of band states and the atomic eigenbasis, which creates $k$-dependent phonon matrix elements from initially constant values in the atomic eigenbasis.

The Mexican-hat shaped electronic band-structure shown in Fig.~\ref{figure0} is modeled using the tight-binding parameters $\epsilon^{p}_{0}=1.95$ eV, $\epsilon^{d}_{0}=-1.95$ eV for the on-site energies, and $V_{\text{pp}}=-V_{\text{dd}}=-0.5$ eV, $V_{\text{pd}}=0.05$ eV for the coupling elements. For the coherent phonons, we take $\hbar \omega_{\bvec{0}} = 12.4$ meV and $\gamma^{P}_{\text{deph}} = 5.0$ ps$^{-1}$. The electron-phonon matrix element controls the influence of the optical phonon on the band dynamics and plays an import role in our model. To study its influence, we present calculations using the values of $g^{pd}_{0}=8.0$ meV and $g^{pd}_{0}=10.0$ meV in the orbital basis for this matrix element. The  optical excitation is patterned after the experimental conditions in Ref.~\onlinecite{Mathias2016} and taken to be a Gaussian pulse with 1.6 eV photon energy, temporal width of $\sigma_T =14$\,fs 
%and energy spread of $\sigma_E = 175$\,meV 
assuming a Rabi energy $\hbar\Omega_{0}=10$\,meV. This results in an excitation of electrons in the conduction band around 200 meV above the unexcited Fermi surface.
 
 After analyzing the consequences of the $k$-dependent basis transformation we will introduce a further simplified model which assumes an averaged value of the electron-phonon matrix element and neglect the influence of the basis transformation. In this case we treat the phonon matrix element as a parameter with $\tilde g^{\mathrm{cc}}_{0}=8.0$\,meV.
 
\subsubsection{Band and carrier response after optical excitation}

The essential characteristics of the dynamical results are discussed for the setup with $g^{pd}_{0}=10.0$~meV, because effects of band renormalization and carrier multiplication are clearly visible in this case. In Fig.~\ref{figure1} snapshots of the band and carrier dynamics of the conduction band are shown. Before the optical excitation the valence band is full and the conduction band nearly empty. The band gap of the Mexican-hat shaped bands is 100 meV at the crease of the Mexican hat $k_0\simeq 0.35\,\text{nm}^{-1}$. At around 0 fs the ultrafast optical pulse excites carriers from the lower lying $\mathrm{v}'$ band into the conduction band c at around 200 meV above the unexcited Fermi energy cf.~Fig.~\ref{figure0}. Between -25 fs and 25 fs mainly optical excitation occurs, but also carrier scattering and the onset of impact ionization. The combination of these effects and the mexican-hat band structure lead to a small second peak at the band bottom $k_0$. Due to the comparatively large band gap of 100 meV, the impact ionization initially is not very efficient. However, after 25 fs, i.e., after the optical excitation is over, the hot carriers in the conduction band lead to a gap closing due to the coherent phonon dynamics and a more efficient screening. The gap closing leads to a more efficient impact ionization, as will be discussed in detail in connection with Fig.~\ref{figure4}. Fig.~\ref{figure4} supports the following scenario: The hot carriers relax from the first peak induced by the optical excitation via impact ionization into the second peak at the band bottom. The impact ionization induces a carrier multiplication in the conduction band, which results in a further gap closing. The smaller gap makes impact ionization even more efficient, which speeds-up the relaxation of the hot carriers, which is visible in the distances between the snapshots in Fig.~\ref{figure1}, but more clearly in Fig.~\ref{figure4} below. Thus there is a mutual amplification between gap closing and impact ionization. The latter occurs predominantly at the band bottom $k_0$ of the mexican hat and thus feeds the second peak on the band bottom of the conduction band in Figure~\ref{figure1} until no more phase space for electron-electron scattering is available and a quasi-equilibrium distribution is reached. 
 
We next investigate details of the carrier and band-gap dynamics for the same parameters as in Fig.~\ref{figure1}, which are marked by solid black lines in Figs.~\ref{figure3} and \ref{figure4}. We defer a discussion of the different parameters (dashed and red curves in Figs.~\ref{figure3} and \ref{figure4}) to the next subsection. The solid black lines in Fig.~\ref{figure3} shows the carrier distribution of the conduction and valence band 250 fs after the optical excitation. As the optical excitation is into the conduction band, the increase of the hole density around the top of the valence band $k_0$ in the first 250 fs after the optical excitation indicates the effect of impact ionization as all the carrier dynamics is exclusively due to Coulomb scattering. In Figure~\ref{figure4} the solid black lines depict the time-dependence of the conduction-band carrier density and the band gap. The fast increase of the carrier density due to the ultrafast optical excitation occurs around the center of the pulse at 0\,fs. This induces a gap closing via coherent phonons that is clearly visible for times later than 50 fs. Finally, and importantly, there is a delayed increase of the carrier density that is exclusively due to impact ionization from carriers originating from the valence band. This impact ionization therefore effectively acts a carrier excitation mechanism which drives the distributions in the conduction and valence bands further away from equilibrium. The coupling of the nonequilibrium carriers to the coherent phonon increases the band-gap shrinkage further.

\subsubsection{Influence of electron-phonon matrix elements on response}

\begin{figure}[tb]
\centering
\includegraphics[scale=0.6]{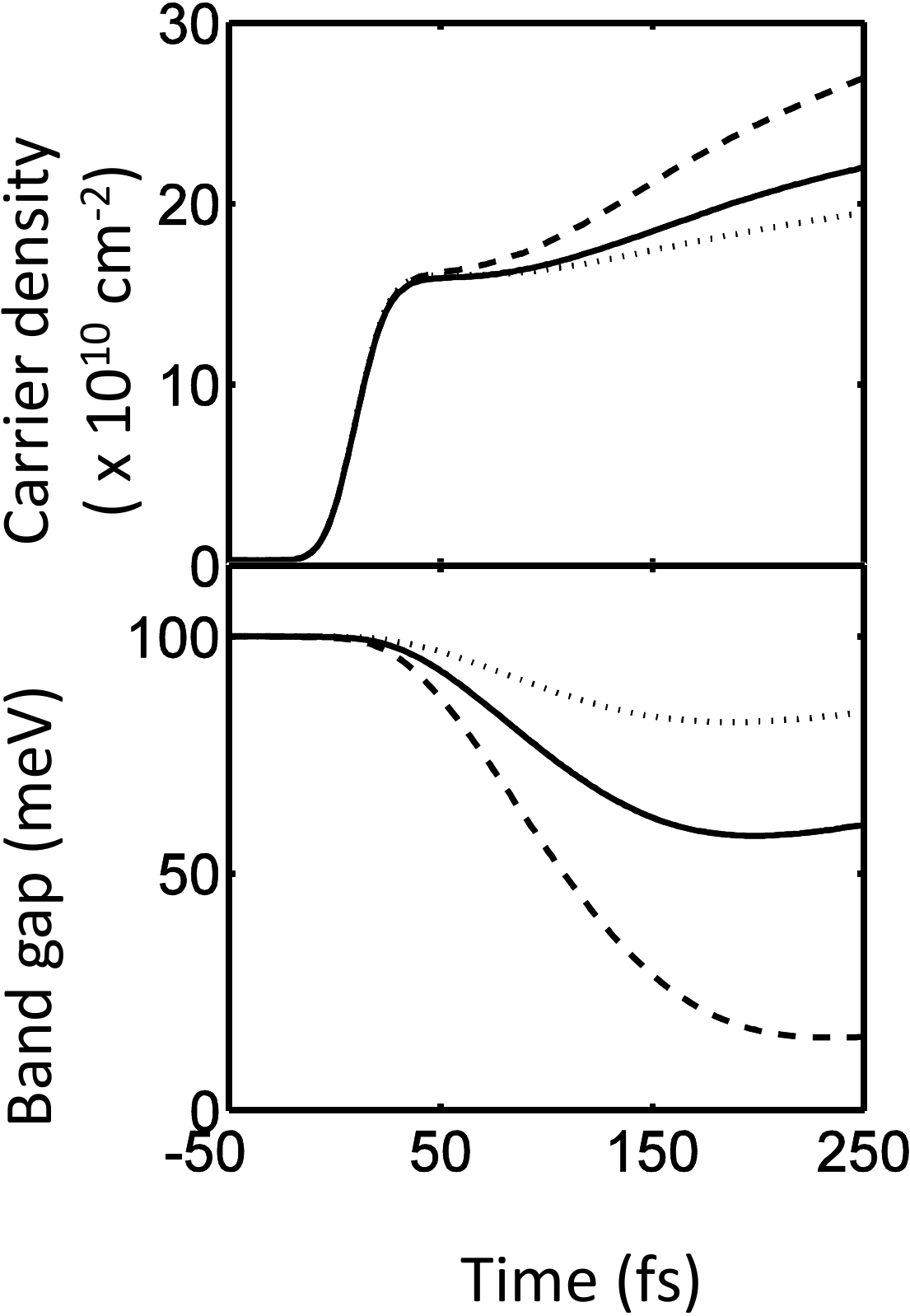}
\caption{Gap closing (bottom) and carrier density in the conduction band (top) vs. time for setups with averaged phonon matrix elements of 4.0~meV (dotted), 6.0~meV (solid) and 8.0~meV (dashed).}
\label{figure5}
\end{figure}

In Figure~\ref{figure3} and Figure~\ref{figure4} we study the influence of the electron-phonon coupling and also the consequence of using and an averaged value of the phonon matrix element $\tilde{g}_{0}$. We first replace the electron-phonon matrix elements $g^{pd}_{0}=10$~meV used so far by an averaged matrix element $\tilde g^{\mathrm{cc}}_{0}=8.0$\,meV. With this replacement, we obtain similar final carrier distributions after 250 fs as shown in Figure~\ref{figure3}, similar carrier densities and band-gap dynamics as shown in Fig.~\ref{figure4} and thus also a similar carrier multiplication of slightly above 70\% at 250\,fs. To show the sensitivity of the results on the electron-phonon coupling matrix element, we also show a calculation with $g^{pd}_{0}=8.0$~meV. This leads to a sizable difference in the final carrier distributions, reduces the band-gap shrinkage and also the carrier multiplication to about 50 \% for the setup with $g^{pd}_{0}=8.0$\,meV. Therefore, for a simulation of real materials and their electron-phonon matrix elements it is important to take into account the basis transformation. However, in the spirit of our model, we will use averaged electron-phonon matrix elements, which are capable of reproducing the dynamical calculations, albeit for a slightly different value of the electron-phonon matrix elements. This is sufficient for the more qualitative analysis of the present paper. 

In Fig.~\ref{figure5} we analyze the influence of different values of the averaged electron-phonon coupling matrix element $\tilde{g}_{0}$ by comparing $\tilde g^{\mathrm{cc}}_{0}=8.0$\,meV, 4.0 meV, and 6.0 meV. Because the quenching of the insulator phase is due to the coherent phonon dynamics, the gap closing depends on the strength of the electron-phonon matrix. Due to efficiency of the coupling between carrier and band dynamics, the mutual amplification between impact ionization and gap closing leads to a gap minimum of 15 meV for the setup with $\tilde g^{\mathrm{cc}}_{0}=8.0$\,meV, 58 meV for the setup with $\tilde g^{\mathrm{cc}}_{0}=6.0$\,meV and 81 meV for the setup with $\tilde g^{\mathrm{cc}}_{0}=4.0$\,meV. Besides the gap closing also the carrier multiplication is visible in Fig.~\ref{figure5} via the time-evolution of the carrier density in the conduction band. After 300 fs a carrier multiplication of 74 \% ($\tilde g^{\mathrm{cc}}_{0}=8.0$\,meV), 42 \% ($\tilde g^{\mathrm{cc}}_{0}=6.0$\,meV), and 26 \% ($\tilde g^{\mathrm{cc}}_{0}=4.0$\,meV) is reached.  

\subsection{Comparison to experimental results and influence of different excitation scenarios\label{expres}}

\begin{figure}[tb]
\centering
\includegraphics[scale=0.6]{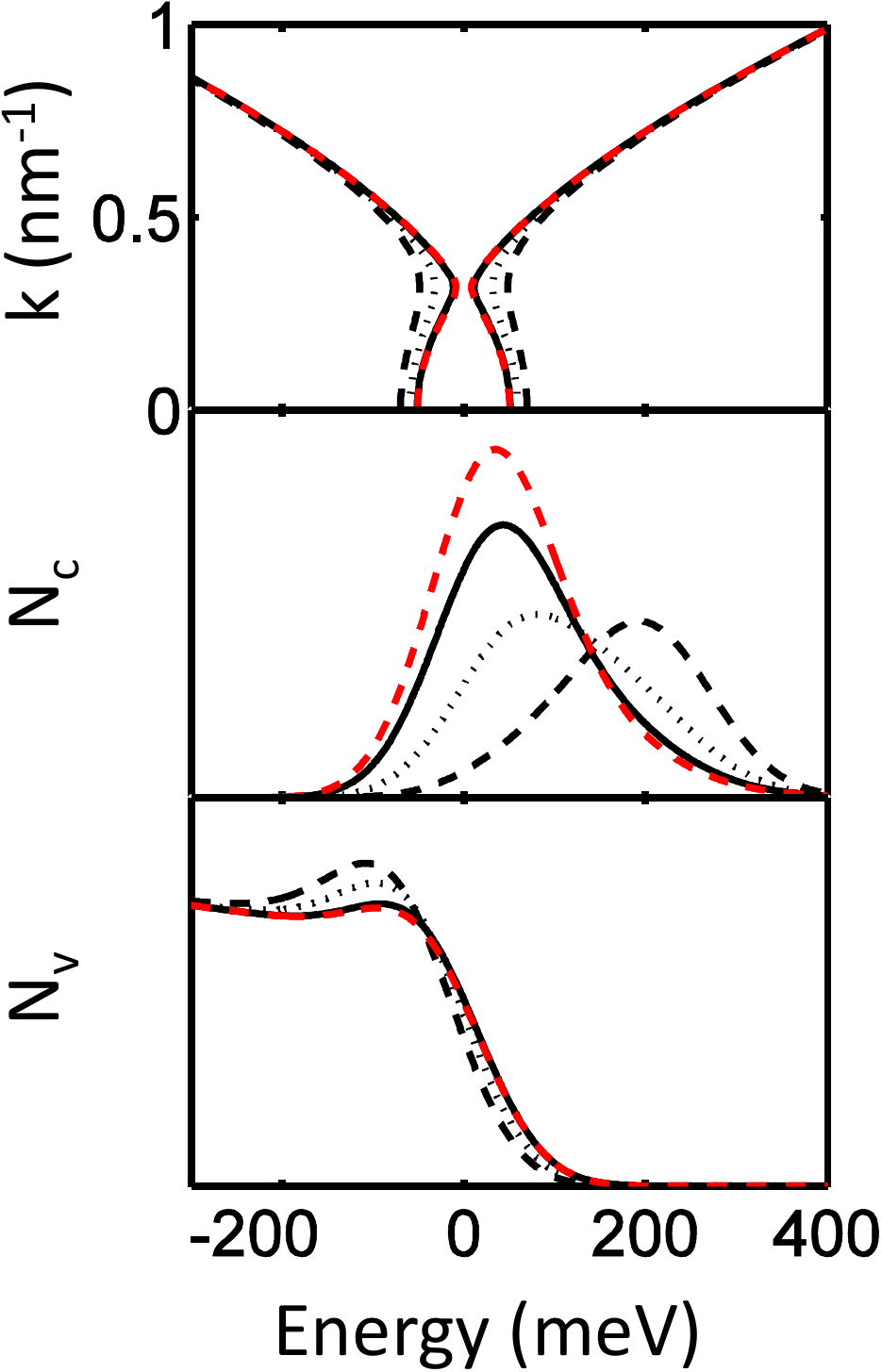}
\caption{Time-dependent band dispersion (top), broadened carrier distribution function in the conduction band (middle) and valence band (bottom) at 25 fs (dashed black), 100 fs (dotted black), 175 fs (solid black), 250 fs (dashed red) after an ultrashort optical excitation at 0 fs. After 200 fs a quasi-equilibrium is almost reached in both bands.   The dynamics of the distribution function in the valence band is mainly due to the gap closing and reshaping of the bands as the effect of impact ionization is hardly visible due to the broadening. Note that for a better comparison between the $N_c$ and $N_v$ curves, $N_v$ has been multiplied by a factor of 16.}
\label{figure6}
\end{figure}

\begin{figure}[tb]
\centering
\includegraphics[scale=0.6]{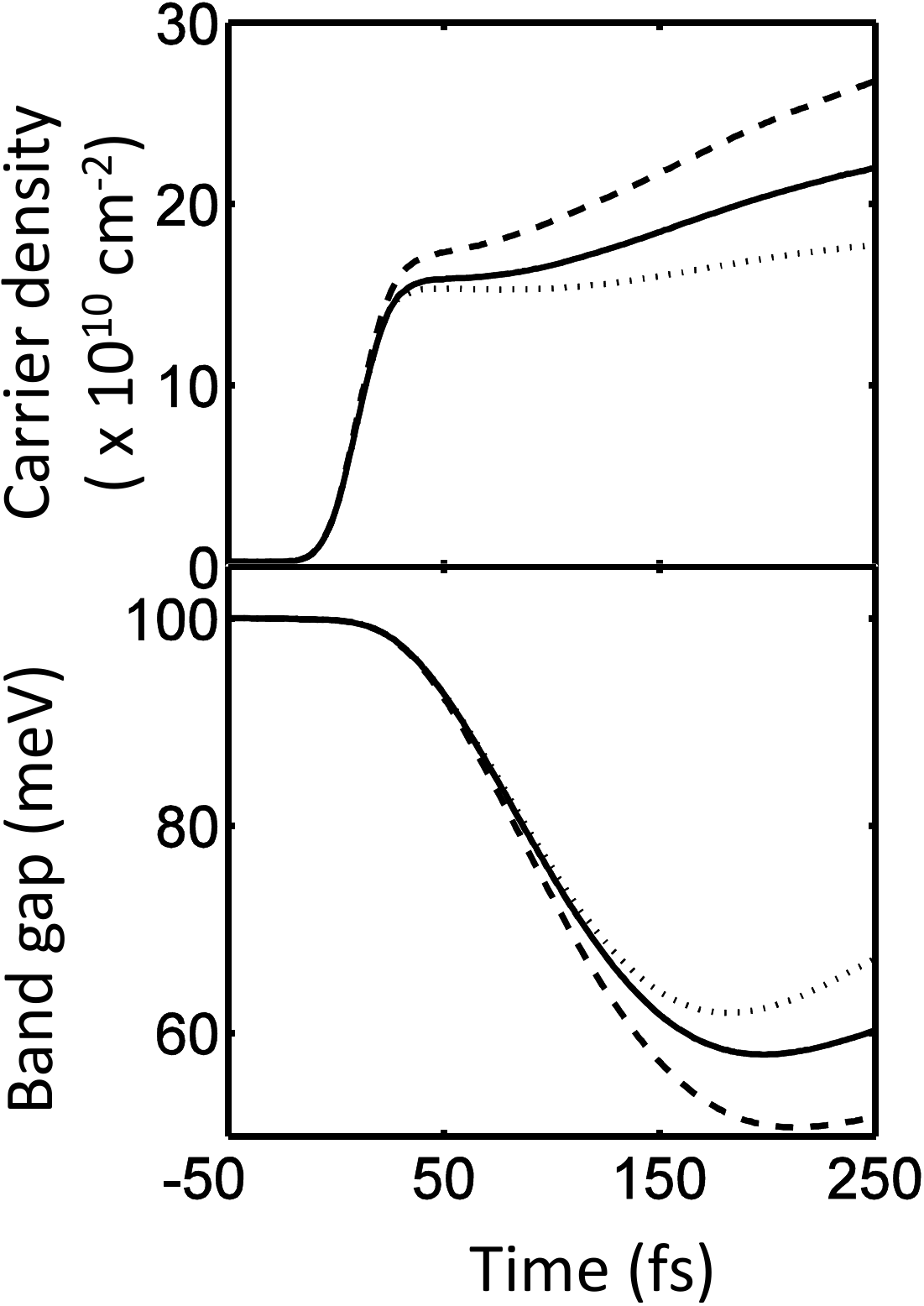}
\caption{Gap closing (bottom) and carrier density in the conduction band (top) vs.\ time for different excitation energies of 150 (dotted), 200 (solid) and 250 (dashed) meV above the Fermi energy.}
\label{figure7}
\end{figure}

\begin{figure}[tb]
\centering
\includegraphics[scale=0.6]{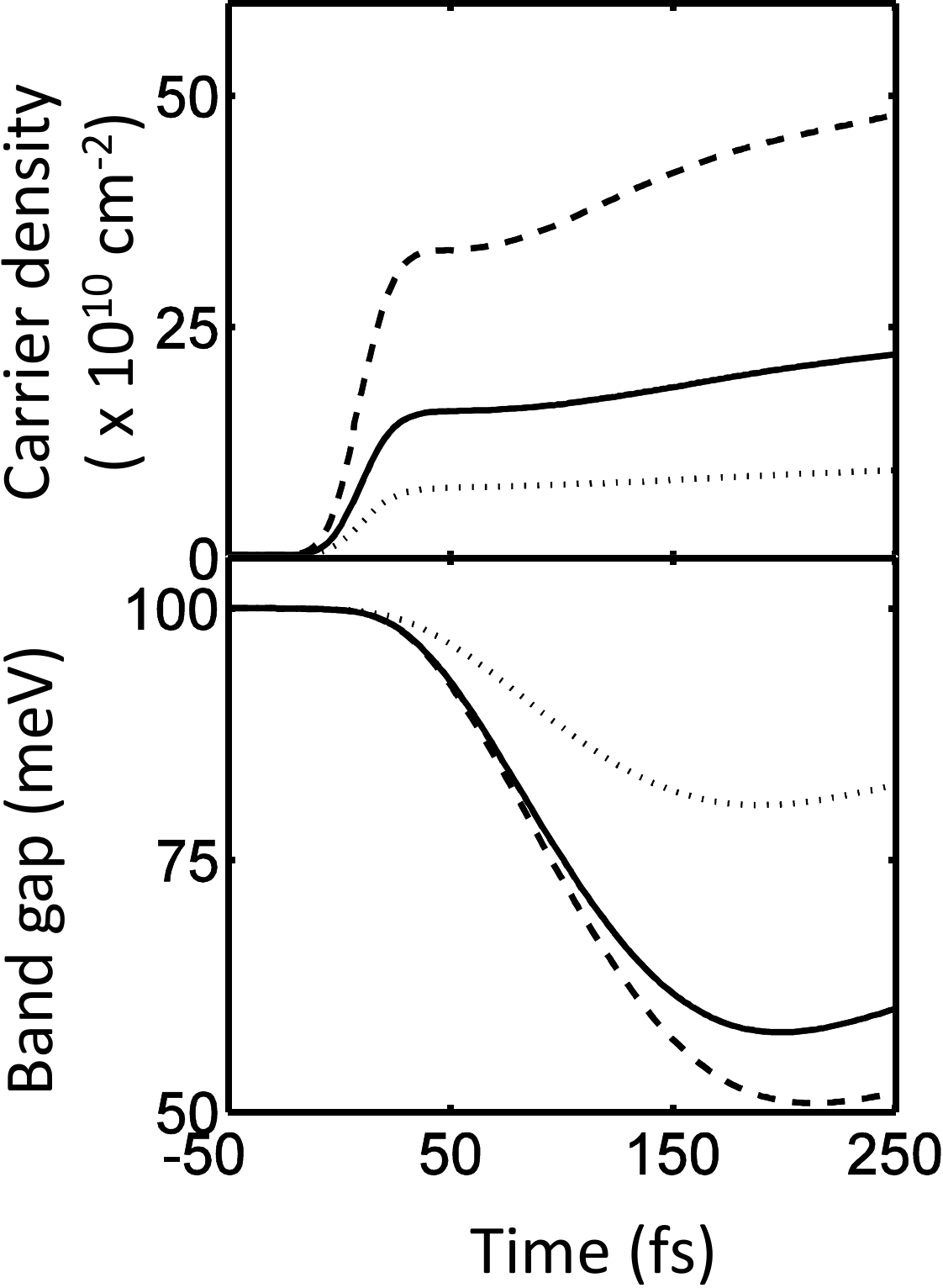}
\caption{Gap closing (bottom) and carrier density in the conduction band (top) vs. time for different Rabi energies of 6.6 (dotted), 10.0 (solid) and 15.0 (dashed) meV.}
\label{figure8}
\end{figure}
An important objective of this paper is to provide results that can be compared with recent experimental photoemission data. In particular, the energy distribution curves of photoemitted electrons cannot unambiguously be interpreted without some theoretical model.~\cite{Mathias2016} As we cannot compute the cross sections that would be needed for a quantitative comparison with the energy distribution curves, we present a qualitative comparison using a broadening of the conduction and valence distribution by the typical experimental energy resolution of 150 meV (FWHM). The broadened distributions are defined by 
\begin{equation}
 N_{b}(E) = \sum_{\bvec{k}} n^{b}_{\bvec{k}} 
 			\ g_{\Delta E} (\epsilon^{b}_{\bvec{k}} - E )
 \label{eq:N-g}
\end{equation}
where $g_{\Delta E} (\epsilon^{b}_{\bvec{k}} - E )$ is a Gaussian of width $\Delta E$. The important point for the comparison with experiment are the energy dependent features, not the numerical value of the $N_{b}$s.

In the following, we first illustrate the spectral and kinetic response with the help of the broadened distributions using our two band model and the electron-phonon matrix element $g^{pd}_{0}=10.0$\,meV in the orbital basis including the basis transformation of the electron-phonon matrix elements. We first show results for our model using the band-gap of TiSe$_2$ in the charge-density wave phase which are intended to be compared to electron distribution curves for small band-gap materials. Afterwards we analyze the dependence of the response on the optical excitation by a parameter study for different Rabi frequencies and excitation energies.

The broadened distributions for conduction and valence bands, together with the band dispersions are shown in Figure~\ref{figure6} for different times. We focus on the distribution of conduction electrons first. In the first 25 fs, the ultrafast optical excitation creates a peak in this distribution function at 200 meV above the Fermi energy. After 100\,fs the hot-carrier relaxation due to impact ionization induced by the gap closing is clearly visible in the broadened distribution function. However, in a  Mexican-hat shaped band-structure the interpretation of the broadened electron distribution is not straightforward because a time-dependent increase in the conduction band is a combination of band-dispersion effects and the on-going carrier multiplication. As we have seen in Figure~\ref{figure3} and Figure~\ref{figure4} the effects of carrier multiplication dominate the increase of the carrier signal at later times. The mutual amplification between impact ionization and gap closing, which has already been discussed, continues as shown by the snapshots for the band and distribution functions in Figure~\ref{figure6} until there is no more phase space for electron-electron scattering available and a quasi-equilibrium distribution is reached after 250 fs. Starting from a value of the band-gap that is realistic for small-bandgap material like 1$T$-TiSe$_2$ we thus obtain in our model calculation a signal of ultrafast carrier dynamics that is in agreement with experimental results, such as those reported in Ref.~\onlinecite{Mathias2016}. Our calculated ``signal'' can be explained in terms of a mutual amplification between gap closing, which goes along with the quenching of the insulator phase, and impact ionization.  In the present paper, the band gap dynamics are due to coherent phonons and are therefore applicable to a Peierls-like insulator. It is to be expected that a similar connection between gap closing and impact ionization occurs also for an excitonic-insulator phase-change mechanism. It may be even more pronounced in the excitonic-insulator case, because there the characteristic response times are faster than the response time of a Peierls insulator, cf.~Ref.~\onlinecite{Hellmann2012}, which indicates that the important electron-electron coupling matrix elements in that case are larger. However, because of the slower gap response in the Peierls (electron-phonon coupling) case, the connection between gap closing and carrier multiplication can be more easily disentangled in the model used here. 

Turning to the broadened valence band distributions, shown in Fig.~\ref{figure6} (bottom), the holes at the top of the Mexican-hat shaped valence band created by impact ionization are hardly visible. This is because the broadening of the distribution function almost completely removes the dip in the microscopic valence band distributions $n_v$ shown in Fig.~\ref{figure3}. This is in agreement with our earlier study using a parabolic valence-band and experimental results in Ref.~\onlinecite{Mathias2016}. The dynamics of the Mexican hat introduces new features, such as the bump of the broadened valence band distribution around the top of the filled valence band, which is due to the changing band dispersion. The shift of the broadened distribution function is due to the band shift of the valence band as shown in Fig.~\ref{figure6} (top). These results give a simple microscopic picture of electronic dynamics underlying the electron distribution curves observed, e.g, in Ref.~\onlinecite{Mathias2016} for TiSe$_2$. We plot and discuss the band and carrier dynamics in this paper up to about 250\,fs, which is the onset of the phase transition. At longer times the system can go further into a different phase where the back-folded valence band disappears, or the system can return to the insulator phase by cooling processes due to carrier-phonon scattering. Both effects are not included in the present study and are left for future investigations. 

After analyzing the characteristic response of the system, we study its dependence on the optical excitation by varying the Rabi frequency and photon energy of the ultrafast optical excitation. The quantitative differences between a calculation with and without the electron-phonon basis transformation are insignificant for this analysis and for the sake of simplicity, we use an averaged electron-phonon matrix element $\tilde{g}^{\mathrm{cc}}_{0}=6.0$\,meV. We take this as a reference value in the following parameter study as it is an intermediate value of the coupling so that a reduction and an increase still show an interesting gap closing dynamics. The value of $\tilde{g}^{\mathrm{cc}}_{0}=10.0$ was chosen in Fig.~\ref{figure6} because it exhibits a relatively fast dynamics (of the gap closing and the carrier dynamics) for which structures in the distribution function are most easily visible.

In Fig.~\ref{figure7} we compare different excitation photon energies, which lead to different energies at which the electrons are created in the conduction band. We call this the excitation energy~$E_X$ and measure it from the unexcited Fermi energy~$E_{\text{F}}$, as sketched in Fig.~\ref{figure1}. We analyze the cases of $E_X=150$\,meV, 200 meV (which has been used so far and constitutes our reference setup in the following) and 250\,meV. For the setup with $E_{X}-E_{F}=250$\,meV, carriers are excited around 200\,meV above the conduction-band bottom; the distance to the band bottom is reduced to 100 meV for the setup with $E_{X}-E_{F}=150$\,meV. As shown in Fig.~\ref{figure7}(top) the carrier density created during the optical pulse in the conduction band is only slightly different for the three setups, but its subsequent time evolution is different.  
 However, the corresponding band gap changes for the different excitation energies in Fig.~\ref{figure7}(bottom) deviate only by around 20\,meV, i.e., 10 meV for each band, which is much smaller than the difference of the excitation energies. The most important contribution to the difference in carrier densities for the three excitation energies must therefore be due to different carrier multiplication effects, and the impact ionization is most efficient for the setup with $E_{X}-E_{F}=250$\,meV. Fig.~\ref{figure7} further shows that the efficiency of the mutual amplification between impact ionization and gap closing increases nonlinearly. After 250\,fs the values for the the carrier multiplication are 73\%, 42\%  and 14\%, respectively, for the excitation energies of $E_{X}-E_{F}=250$\,meV, 200\,meV and 150\,meV. The corresponding band gaps are 52 meV, 60 meV, and 67\,meV. This nonlinearity is mainly due to repeated interband scattering processes that become possible for electrons excited at higher energies. During their scattering dynamics toward the band bottom these electrons can contribute to the carrier multiplication process twice or more times. 

In Fig.~\ref{figure8} we investigate the dependence of the dynamics on the excitation strength. We compare three amplitudes of the Rabi energy $\hbar \Omega_0$: 6.6\,meV, 10.0\,meV (our reference setup and the value used so far) and 15.0\,meV. In difference to Fig.~\ref{figure7}, the carrier density created in the conduction band by the optical excitation is different for the three cases. The carrier density is the driving force of the coherent-phonon amplitude, which induces an atomic displacement responsible for the band gap dynamics, so that we obtain a higher initial band gap reduction for larger values of $\hbar\Omega_{0}$ and this band gap remains smaller due to the mutual amplification between gap closing and impact ionization. The gap closing evidently saturates in Fig.~\ref{figure8}(bottom). The effect of impact ionization can be assessed from the carrier multiplication in Fig.~\ref{figure8}(top), which is 33\%, 42\% and 43\%, respectively for Rabi energies $\hbar\Omega_{0}=6.6$\,meV, 10.0\,meV and 15\,meV. This carrier multiplication shows only a comparatively small increase between the two smaller Rabi energies whereas we have a pronounced difference in the gap closing. The difference between the dynamical scenarios shown in Fig.~\ref{figure8} is therefore mainly due to the different gap closing related to the initial photoexcited carrier density and the saturation of the gap closing is mainly responsible for the saturation of the carrier multiplication. Parenthetically we remark that a saturation of the gap quenching of the charge-density wave state has been observed in the charge-density wave material~RTe$_3$ where only an incomplete suppression of the charge-density wave occurs; here, we find an indication of a saturation for comparatively small electron-phonon coupling.~\cite{Rettig2016:NatComm}

\subsection{Influence of different band shapes}

\begin{figure}[tb]
\centering
\includegraphics[scale=0.6]{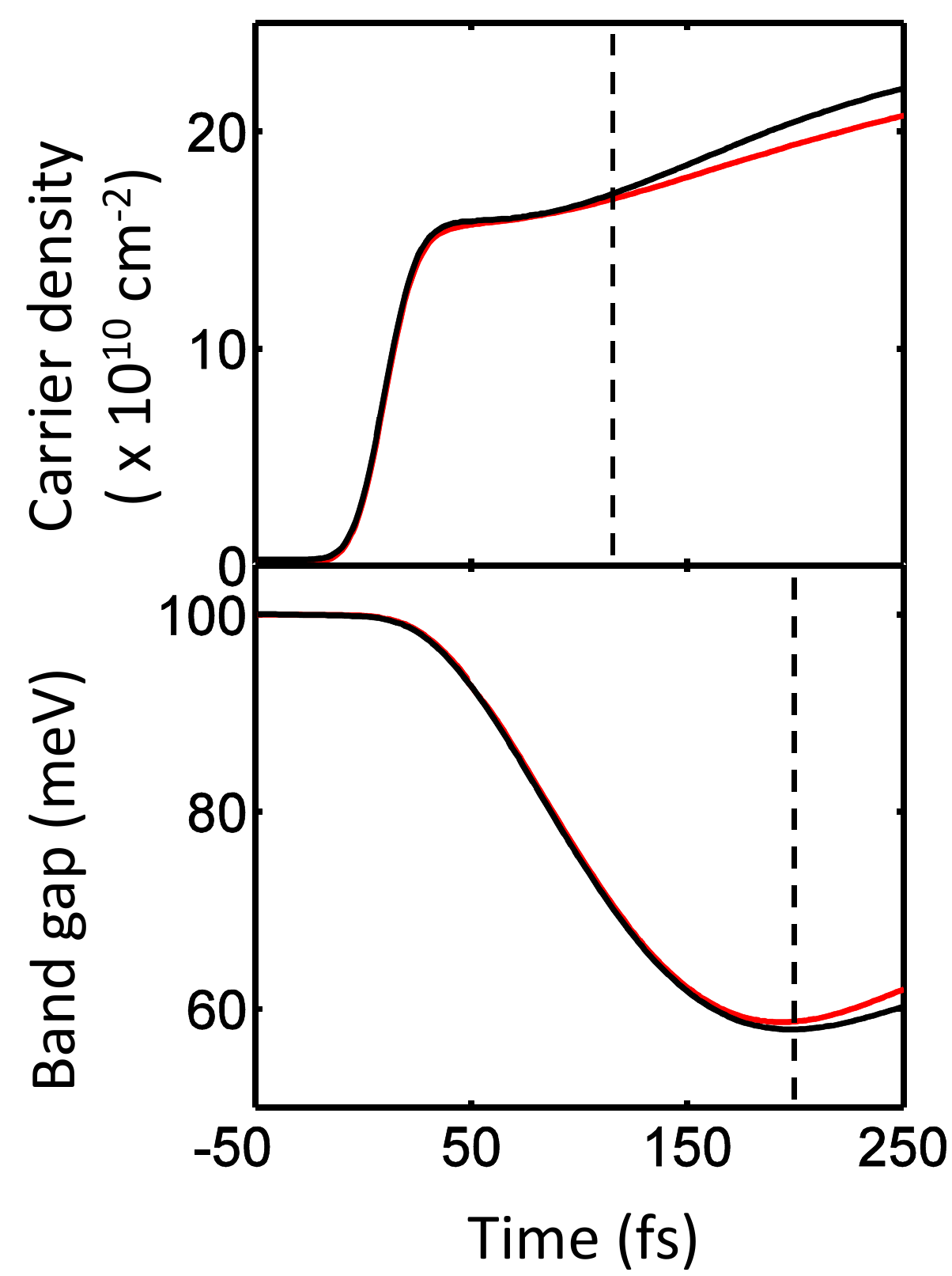}
\caption{Gap closing (bottom) and carrier density in the conduction band (top) vs.~time for a parabolic (red) and a Mexican-hat shaped (black) band with phonon matrix element $\tilde g^{\mathrm{cc}}_{0}=6.0$\,meV.}
%\label{figure11}
\label{figure9}
\end{figure}

\begin{figure}[tb]
\centering
\includegraphics[scale=0.6]{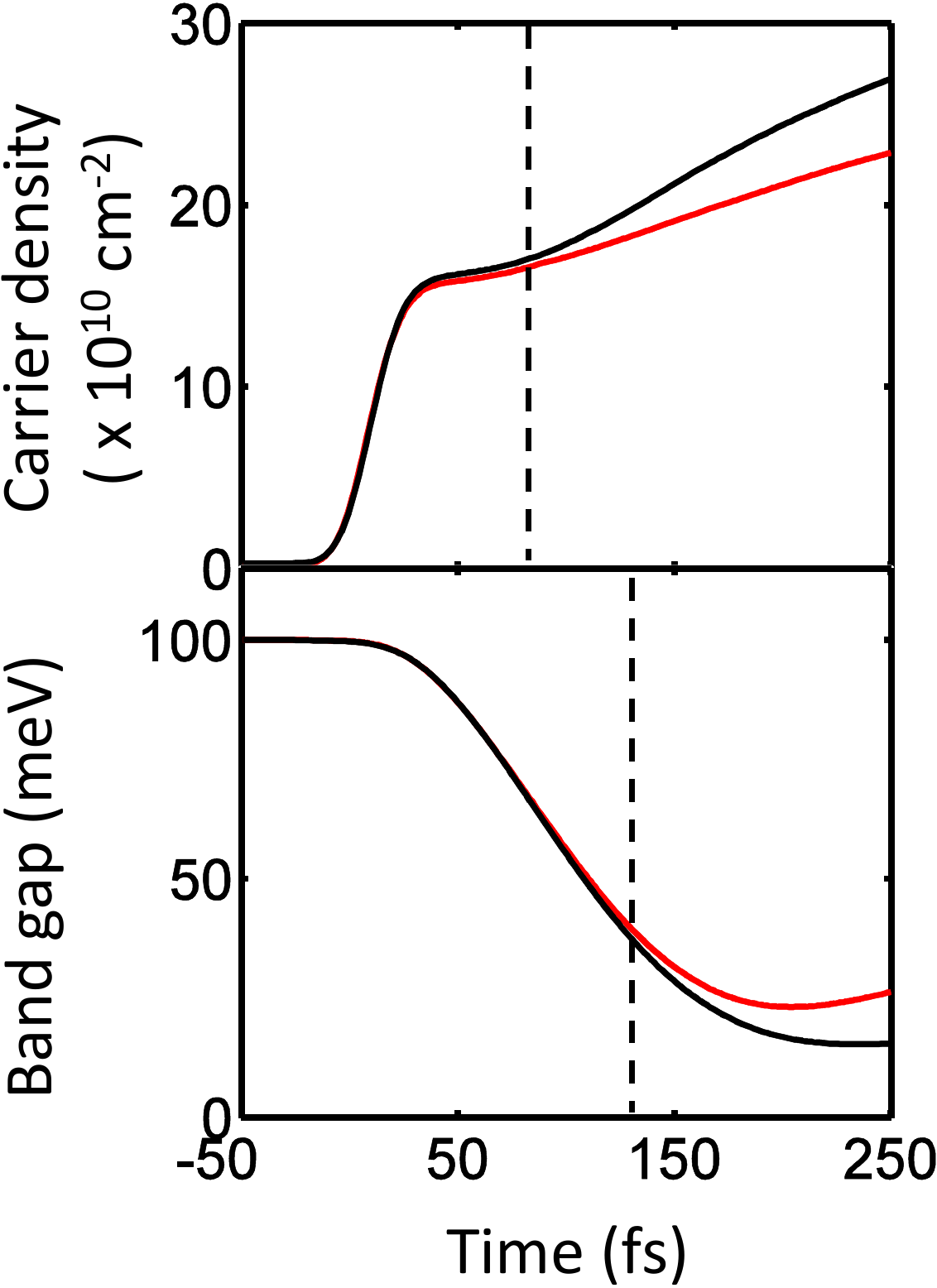}
\caption{Gap closing (bottom) and carrier density in the conduction band (top) vs.~time for a parabolic (red) and a Mexican-hat shaped (black) band with phonon matrix element $\tilde{g}^{\mathrm{cc}}_{0}=8.0$\,meV.}
%\label{figure11b}
\label{figure10}
\end{figure}

\begin{figure}[tb]
\centering
\includegraphics[scale=0.6]{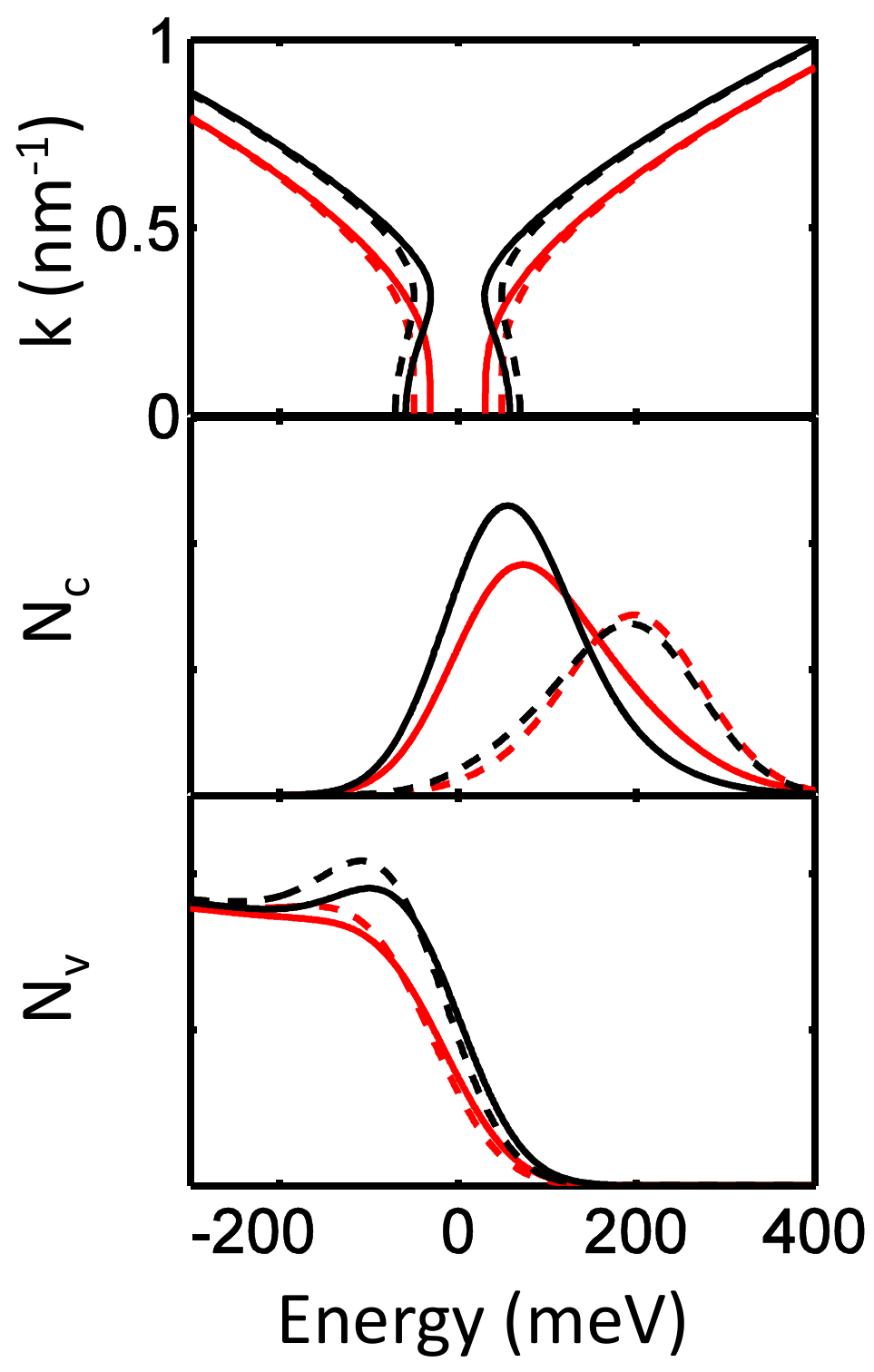}
\caption{Band dynamics (top) and broadened carrier distribution function in the conduction band (middle) and valence band (bottom) at 25 fs (dashed) and 250 fs (solid) after an ultrashort optical excitation at 0 fs for a parabolic (red) and a Mexican-hat shaped (black) band with phonon matrix element $\tilde{g}^{\mathrm{cc}}_{0}=6.0$\,meV.}
%\label{figure12}
\label{figure11}
\end{figure}

\begin{figure}[tb]
\centering
\includegraphics[scale=0.6]{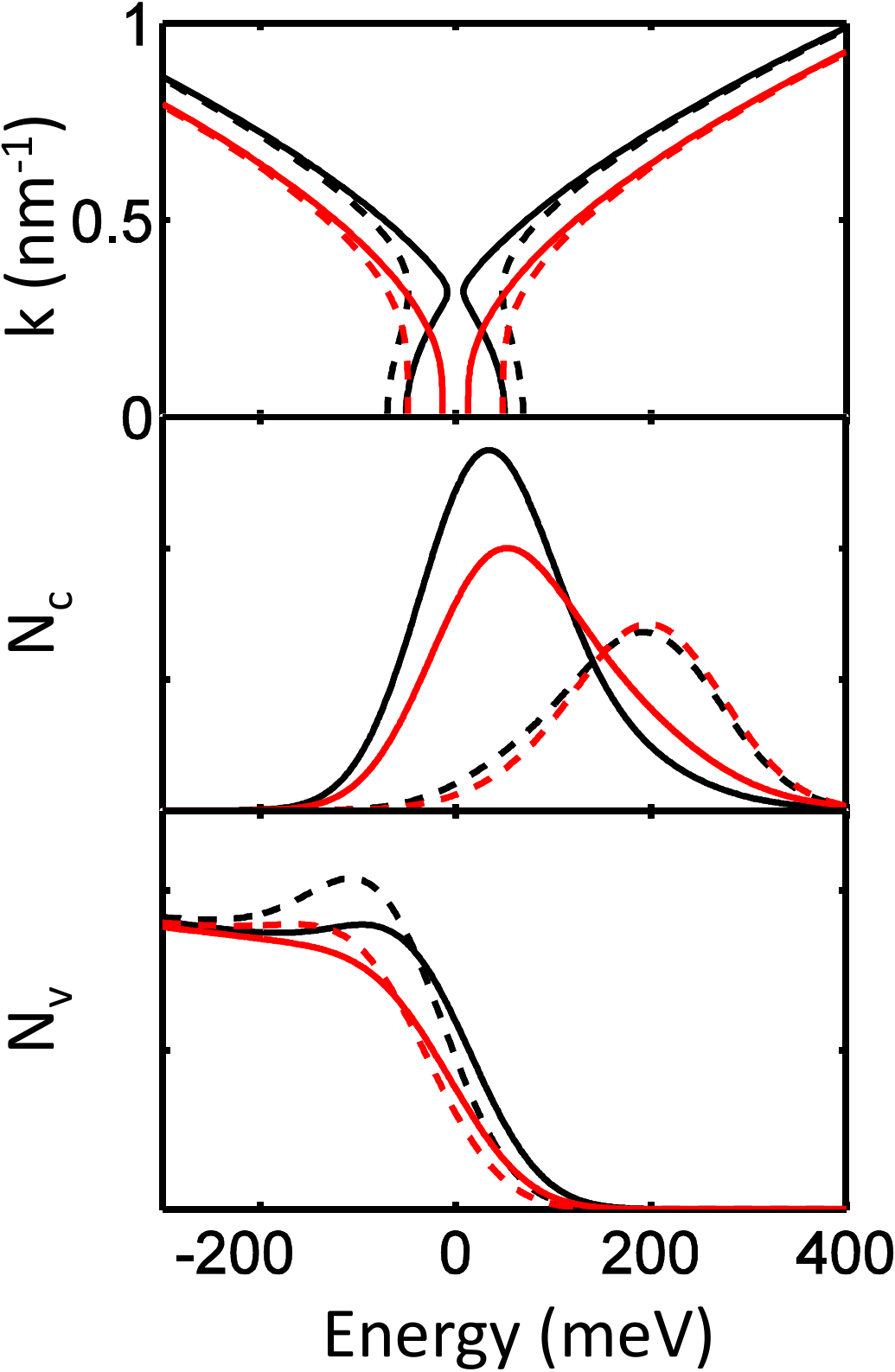}
\caption{Band dynamics (top) and broadened carrier distribution function in the conduction band (middle) and valence band (bottom) at 25 fs (dashed) and 250 fs (solid) after an ultrashort optical excitation at 0 fs for a parabolic (red) and a Mexican-hat shaped (black) band with phonon matrix element $\tilde{g}^{\mathrm{cc}}_{0}=8.0$\,meV.}
%\label{figure12b}
\label{figure12}
\end{figure}

As already mentioned, we are interested in elucidating the influence of the band structure on measurable quantities, in particular energy distribution curves produced by photoemission experiments. In order to understand the calculated broadened distribution functions that can be compared with experiment, we here first discuss the influence of the band shape on the carrier dynamics without added broadening, and use for the comparison a parabolic band and the Mexican hat shaped band that we have based our calculations on so far. For a meaningful comparison, we define the parabolic band setup using all band parameters of the Mexican-hat like band setup, except a change of the on-site energies $\epsilon^{p}_{0}$ and $\epsilon^{d}_{0}$ from $1.95$\,eV to $2.0$\,eV. In this way, the parabolic and the Mexican-hat shaped bands have the same band gap, but in the parabolic case it occurs at $k=0$ and in the Mexican-hat band case at $k_0$. The band structures are plotted in Figs.~\ref{figure11}(top) and \ref{figure12}(top) as dotted lines.

In Fig.~\ref{figure9} and Fig.~\ref{figure10} the gap closing and carrier density in the conduction band vs.~time is shown for the parabolic and Mexican-hat shaped band structures with $\tilde g^{\mathrm{cc}}_{0}=6.0$\,meV and $\tilde g^{\mathrm{cc}}_{0}=8.0$\,meV, respectively. Due to the different band shape, the ''tuning'' of the band dispersion and the optical excitation is slightly different. We have already seen in Fig.~\ref{figure7} that such a small difference in excitation energy will also lead to a slightly different optically excited carrier density for the two band structures. However, the further time evolution of the carrier density is mainly determined by the different band dispersions. The origin of the steeper band dispersion for the Mexican-hat shaped band is the characteristic band gap minimum at a $k_0\neq 0$ (i.e., not at the high-symmetry point), and a local band gap maximum at $k=0$ (the high symmetry point) in contrast to the parabolic case, where the global band gap minimum is at $k=0$. The efficiency of the impact ionization depends on the size of the band gap and the $k$-dependent Bloch wavefunctions, which include band mixing effects, as well as the available phase space for the scattering process. The band mixing is connected to the position of the band gap minimum and, thus, different between the parabolic and Mexican-hat setups.

These qualitative differences between the band structures should lead to different carrier dynamics, and we compare these dynamics in Figs.~\ref{figure9}--\ref{figure12}. In a parabolic band, the hot carriers relax directly into the high symmetry $k=0$ point, while in the case of the Mexican-hat shaped band, hot carriers relax more into the band minimum $k_0$ and reach the local maximum at the high symmetry $k =0$ point only with a delay. Therefore, band gap minima on different $k$ positions combined with a different $k$ dependence of the available phase space results in different efficiency for impact ionization for equal band gap minima. We first focus on Figs.~\ref{figure9} and Fig.~\ref{figure10} where we compare the gap closing and conduction-band carrier density between Mexican-hat and parabolic bands for different electron-phonon couplings. A higher impact ionization efficiency for the Mexican-hat shaped band induces a difference in the carrier density between the two band structures, as can be seen from splitting of the curves above 100 fs in Figs.~\ref{figure9} and Fig.~\ref{figure10}, respectively. The mutual amplification between impact ionization and band gap closing amplifies the difference in the temporal evolution of the band gap (see splitting of the curves at a slightly later time around 150 fs) and of the impact ionization efficiency. Therefore, the difference between the two band structures increases for band gap and conduction-band carrier density with time. In Fig.~\ref{figure9} this results in a band gap of 60\,meV and 62\,meV and a carrier multiplication of 42\% and 35\% (factor: 1.2) for the Mexican-hat structures compared to the respective calculation with the parabolic bands. In Fig.~\ref{figure10} for the larger electron-phonon coupling $\tilde{g}^{\mathrm{cc}}_{0}=8.0$\,meV  we have band gaps of 15\,meV and 26\,meV after 250 fs, and carrier multiplications of 74\% and 50\% (factor: 1.5) for the Mexican-hat and parabolic bands, respectively. Depending on the electron-phonon matrix element the influence of the band shape can therefore be substantial.

In Figs.~\ref{figure11} and~\ref{figure12} we compare the time evolution of the band structure and broadened carrier distributions for the Mexican-hat and parabolic bands. As the broadened distributions average over all $k$ states in a given energy range set by $\Delta E$ in Eq.~\eqref{eq:N-g} they are influenced both by the carrier redistribution dynamics (i.e., carrier multiplication) and by the band structure, especially \emph{if} the band structure changes. In our earlier paper,~\cite{Mathias2016} we presented a simple parabolic model without band mixing and without a dynamically changing band structure. With the present calculation for parabolic and Mexican-hat shaped bands and a consistent inclusion of band mixing effects, we can investigate the contribution of the dynamical band structure to the electron distribution curves.

Fig.~\ref{figure11} and Fig.~\ref{figure12} are for different electron-phonon couplings and are obtained from the same dynamical calculations as Figs.~\ref{figure9} and \ref{figure10}, respectively. For a broadening that corresponds to state-of-the-art photoemission experiments, the broadened carrier distributions created by the optical excitation around 25\,fs are similar for the Mexican hat and parabolic bands, but a difference of the weighted carrier distribution in the conduction band can be seen after 250\,fs. 
In the parabolic-band case, there are only rigid band shifts and negligible changes in band curvature so that the change in the broadened distribution reflects essentially only the redistribution by scattering processes and the increase in the number of carriers due to carrier multiplication. In the Mexican-hat band structure, there are pronounced changes in the band curvature that also influence the broadened distribution functions. While there are quantitative differences to the parabolic case, we observe a similar behavior of the broadened distribution function that is in qualitative agreement with the carrier multiplication factors determined from Figs.~\ref{figure9} and \ref{figure10}. We conclude that the behavior of computed broadened distribution functions, which are the quantity we compare to experimental electron-distribution curves and which are similar to experimental results on TiSe$_2$, are an unambiguous indicator of carrier multiplication. The energy dependent signatures found in Figs.~\ref{figure11} and \ref{figure12} are only influenced to a small extent by changes in the spectral properties of the carriers, even though the  dynamical changes in the curvature of the Mexican hat band structure model likely overestimates those occurring in a real small-band gap material.

In the valence band, a characteristic bump below the Fermi energy appears only in the broadened distribution function in the Mexican-hat shaped band. The erosion of this bump, which is not visible in the parabolic band, indicates the effect of carrier multiplication. In TiSe$_2$ the Se 4p band, as opposed to the Ti 3d, does not show a Mexican hat-like structure and therefore it is difficult to observe characteristics of impact ionization in the valence band with the energy broadening introduced by current experimental photoemission setups. Because the temporal evolution of the valence band signal is more influenced by the band and less influenced by the carrier dynamics, the difference between different band shapes, in particular in Fig.~\ref{figure11}, is more pronounced than the differences between the two snapshots at different times. 

\section{Conclusion}\label{sec:conclusion}

We investigated carrier multiplication dynamics due to impact ionization after ultrafast optical excitation in a model band structure of a quasi-two dimensional material with small band gaps. The photo-induced band gap narrowing close to the Fermi surface is incorporated using the coupling to coherent phonons, which mimics the quenching of an insulator phase. We used a dynamical approach that includes time-dependent band-energies and wave-functions, which make the Coulomb-matrix elements and the static screening effectively time-dependent. Using this model, we were able to quantify the contribution of impact ionization in the ultrafast response of small band-gap 2D materials and discussed the importance of the \emph{interplay} between carrier and band dynamics. We computed broadened distribution functions that can be compared to energy distribution curves as they are measured in time-dependent photoemission spectroscopy, and we discussed the signatures of impact ionization and gap closing in these curves. We also investigated the influence dynamical changes in the band curvature, as these changes will also influence energy distribution curves and cannot, at present experimental resolutions, be distinguished from carrier multiplication effects. To this end, we compared a parabolic band structure with that of a Mexican hat and found that the characteristic change in energy distribution curves in, e.g., TiSe$_2$,~\cite{Mathias2016} are indeed mainly due to carrier multiplication effects, and only to a small extent due to changes in the spectral function of the electrons. Our computed energy dependent distribution curves compare well with experiments on TiSe$_2$ and, even though we consider a specific coupling mechanism to a coherent phonon, we believe that our results capture a general trend in small band-gap 2D materials.

\appendix 

\section{Tight-binding model}\label{app_tb}
%------------------------------------------
For our tight-binding model we assume a quasi-two dimensional material like TMDCs with two kind of atoms A$_{d}$ and A$_{p}$. In the case of a TMDC, atom sort A$_{d}$ would be the transition metal atom (e.g. Ti) with d-type or f-type valence orbital and atom sort A$_{p}$ would be the chalcogen atoms (e.g. Se) with p-type valence orbitals.
The unit cell would consist of one A$_{d}$ and two A$_{p}$ atoms. 
For example the lattice vectors $L_{1}=(l_{1},-l_{2},0)$, $L_{2}=(l_{1},l_{2},0)$ and $L_{2}=(0,0,l_{3})$ would span a unit cell with the atom basis $B_{d}=(0,0,0)$ for A$_{d}$, $B_{p,1}=(b_{1},b_{2},b_{3})$ for the first A$_{p}$ and $B_{p,2}=(b_{1},b_{2},-b_{3})$ for the second A$_{p}$. 
The nearest-neighboring A$_{d}$ or A$_{p}$ atoms in the same plane would have a hexagonal or tetragonal symmetry. To model an accurate bandstructure for a TMDC around the Fermi surface the three $t_{2g}$ (i.e. $d_{xy}$, $d_{zx}$, $d_{xy}$) and eventually the energetically higher two $e_{g}$ (i.e. $d_{3z^2-r^2}$, $d_{x^2-y^2}$) orbitals of the atom of sort A$_{d}$ and the six p-orbitals of the two atoms of sort A$_{p}$ might be considered.\cite{VanWezel2010a} 

Weak interactions between neighboring orbitals are usually neglected and the remaining interactions are expressed in terms of Slater-Koster integrals.\cite{SlaterKoster1954} The bond integrals between two orbitals are distinguished between $\sigma$, $\pi$ or eventually $\delta$ bondings. For example, as described in Ref.~\onlinecite{VanWezel2010,VanWezel2010a} for TiSe$_{2}$, only the three hopping pathways dd$_{\sigma}$, pp$_{\sigma}$ and pd$_{\pi}$ contribute significantly to the behavior of charges close to the Fermi energy.
The resulting band structure around the high symmetry points under investigation of the small band-gap TMDCs is often highly un-isotropic like in TiSe$_{2}$ as reported, e.g., in Ref.~\onlinecite{Monney2010}. For this material, the non-isotropic band dispersion of the Ti 3d band is non-parabolic, i.e., an ellipsoid, and has a Mexican-hat shape geometry in the CDW insulator phase. 
 
Regarding the investigated band dynamics of such a material in the insulator phase, we avoid a material realistic description, where the insulator phase is determined from the normal phase, to investigate the role of different kinds of interactions in the phase transition. Instead we model the TB Hamiltonian already in the insulator phase and describe the band dynamics via an effective atomic displacement as disturbance of the insulator phase. To model the small band-gap around the fermi surface, we use a simple two-band tight-binding model capable to describe the characteristics of the carrier and band-dynamics of such a material after an ultrashort optical excitation. Thus, we obtain an isotropic band-shape around a high symmetry point. The validity of this assumption is additionally motivated at the end of this section.

To transform a more material-realistic TB model into a simpler model with high symmetry capable to describe the characteristics of the carrier and band-dynamics close to the important high-symmetry point, a lot of more or less sophisticated transformation can be done. A simple way of doing it, is to disregard the $t_{2g}$ and to consider only one $d$ orbital, e.g., $d_{xy}$, of A$_{d}$ and one $p$ orbital, e.g., $p_{y}$, of A$_{p}$ and give only $V_{pp\sigma}$, $V_{dp\sigma}$ and $V_{dd\sigma}$ finite values.  In the spirit of such a transformation, we use the following effective tight-binding hamiltonian to describe the investigated small band-gap insulator phase % (to check)
\begin{equation}
\begin{split}
H_{\text{TB}} & = \epsilon^{p}_{0} c^{ps \dag}_{\bvec{k}} c^{ps}_{\bvec{k}} + \epsilon^{c}_{0} (c^{ds}_{\bvec{k}})^{\dagger} c^{ds}_{\bvec{k}} \\ 
& + 2 V_{\text{pp}} \big[ \cos(k_{x} e_{x}) + \cos( k_{y} e_{y}) \big](c^{ps}_{\bvec{k}})^{\dagger} c^{ps}_{\bvec{k}} \\ 
& + V_{\text{pd}} e^{-i \bvec{k}\cdot \bvec{d}_{pd} } (c^{ps}_{\bvec{k}})^{\dagger} c^{ds}_{\bvec{k}}  \\ 
& + V_{\text{dp}} e^{i \bvec{k} \cdot \bvec{d}_{pd} } (c^{ds}_{\bvec{k}})^{\dagger} c^{ps}_{\bvec{k}}  \\ 
& + 2 V_{\text{dd}} \big[ \cos(k_{x} e_{x}) + \cos(k_{y} e_{y}) \big] (c^{ds}_{\bvec{k}})^{\dagger} c^{ds}_{\bvec{k}} 
\end{split}
\label{app_tb1}
\end{equation}
where $\epsilon^{p}_{0}$, $\epsilon^{d}_{0}$ are the on-site energies, $V_{\text{pp}}$, $V_{\text{pd}}=V_{\text{dp}}$, $V_{\text{dd}}$ are the tight-binding coupling-elements and $\bvec{d}_{pd}$ is the relative distance vector between the two effective atoms within the unit cell. As we do not include spin-orbit coupling, we do not explicitly write out the spin-dependence in the following.
The outcome is a conduction band mainly originating from a d-type transition metal orbital and a valence band mainly originating from a p-type chalcogen orbital. In the region of the examined high symmetry point, we obtain a angular symmetric Mexican-hat shaped band with a small band gap and a high band mixing. However, assuming a fast angular redistribution of carrier via electron-phonon scattering as e.g. reported in Ref.~\onlinecite{Mittendorff2014}, the fundamental results of this investigation are also transferable to non-parabolic band structures.

\section{Equation-of-motion in the time-dependent eigenbasis}
\label{app_eom}
%-------------------------------------------------------------
%% Rechnung komplett in n's bezogen auf die zeitabhaengigen Matrixelemente

We start from the total Hamiltonian
\begin{equation}
	H_{\text{tot}}= H_{\text{qp}} + H_{\text{int}}
\label{app_eom1}
\end{equation}
consisting of a quasi-particle hamiltonian $H_{\text{qp}}$ and an interaction hamiltonian $H_{\text{int}}$. If the quasi-particle part is time-dependent, as discussed in section~\ref{sec:coherent-phonons}, where $H_{\text{qp}}= H_{\text{eff}} = H_{\text{CohPh}}+H_{\text{TB}}$, the eigenvalues and eigenvectors of this Hamiltonian have to be calculated for every time-step of the dynamical calculation. Such a time-dependent basis is associated with a basis transformation of the whole equation of motion for every time step as discussed in the following. 

We start using the eigenbasis of quasi-particle hamiltonian $H_{\text{qp}}(t_0)$ at a time~$t_0$. In this basis the equation of motion for the reduced density matrix  $\rho^{b_{1}b_2}_{\bvec{k}} =\langle c^{b_{2}\dag}_{\bvec{k}}c^{b_{1}}_{\bvec{k}} \rangle$ is
\begin{equation}
		\frac{d}{dt} \rho^{b_{1}b_2}_{\bvec{k}}  = \frac{d\rho^{b_{1}b_2}_{\bvec{k}}}{dt}\Big|_{\text{qp}} + \frac{d\rho^{b_{1}b_2}_{\bvec{k}}}{dt}\Big|_{\text{int}}
\label{app_eom2}
\end{equation}
The quasi-particle part of the equation-of-motion can be written as 
\begin{equation}
	\frac{d\rho^{b_{1}b_{2}}_{\bvec{k}}}{dt}\Big|_{\text{qp}} =  h^{b_{1}b_{2}}_{\bvec{k}} \rho^{b_{1}b_{2}}_{\bvec{k}} - h^{b_{2}b_{1}}_{\bvec{k}} \rho^{b_{2}b_{1}}_{\bvec{k}}
\label{app_eom3}
\end{equation}
and the interaction part can be written in the general form of
\begin{equation}
\frac{d\rho^{b_{1}b_2}_{\bvec{k}}}{dt}\Big|_{\text{int}} = \sum \Gamma[\rho]  
\label{app_eom4}
\end{equation}
i.e. a sum over correlation contributions~$\Gamma$, which are functionals of the density matrices~$\rho$.

For a time $t_1>t_0$, the equation of motion $\frac{d}{dt} \tilde{\rho}^{b_{1}b_2}_{\bvec{k}}$ in the new eigenbasis of the system at time $t_1$ takes the form 
\begin{equation}
\begin{split}
	\frac{d}{dt} \tilde{\rho}^{b_{1}b_2}_{\bvec{k}} =& \frac{d}{dt} \left[ U^{\dag} \rho^{b_{1}b_2}_{\bvec{k}} U \right] \\
	 =& \frac{dU^{\dag}}{dt} \rho^{b_{1}b_2}_{\bvec{k}} U 
+ U^{\dag} \frac{d \rho^{b_{1}b_2}_{\bvec{k}}}{dt} U  
+ U^{\dag} \rho^{b_{1}b_2}_{\bvec{k}} \frac{dU}{dt}
\end{split}	
\label{app_eom5}
\end{equation}
where $U$ is a unitary matrix of the basis transformation between the eigenbasis at the time $t_0$ and the eigenbasis at the time $t_1$. For the quasi-particle part we obtain
\begin{equation}
	\begin{split}
		U^{\dag} \frac{d\rho^{b_{1}b_{2}}_{\bvec{k}}}{dt}|_{\text{qp}} U &= U^{\dag} h^{b_{1}b_{2}}_{\bvec{k}} UU^{\dag} \rho^{b_{1}b_{2}}_{\bvec{k}} U \\ 
		& \quad- U^{\dag} T^{b_{2}b_{1}}_{\bvec{k}} UU^{\dag} \rho^{b_{2}b_{1}}_{\bvec{k}} U \\
   &= \tilde{h}^{b_{1}b_{2}}_{\bvec{k}} \tilde{\rho}^{b_{1}b_{2}}_{\bvec{k}} - \tilde{h}^{b_{2}b_{1}}_{\bvec{k}} \tilde{\rho}^{b_{2}b_{1}}_{\bvec{k}}  \\ 
   &= \frac{d\tilde{\rho}^{b_{1}b_{2}}_{\bvec{k}}}{dt}|_{\text{qp}}
	\end{split}
	\label{app_eom6}
\end{equation}

The transformation of the interaction part can be done in an analogous fashion, but the derivation depends on the level of approximation employed for the interaction. We assume here that the interaction part can finally be written in a general form as
\begin{equation}
\frac{d\rho^{b_{1}b_2}_{\bvec{k}}}{dt}\Big|_{\text{int}} = \sum \tilde{\Gamma} [\tilde{\rho}]  
\label{app_eom7}
\end{equation}
One advantage of such a basis transformation is that $\tilde{n}^{b}_{\bvec{k}} = \tilde{\rho}^{bb}_{\bvec{k}}$ can be interpreted as the occupation of the state $|b\bvec{k} \rangle$ at time $t_1$. Therefore, the intuitive physical picture  used in common approximation schemes is preserved. However, the basis transformation is associated with a transformation of diagonal density-contributions $n^{b}_{\bvec{k}}$ into off-diagonal coherence-contributions in conjunction with correlated correction-terms in the equation-of-motion as shown above.

In the case of a sufficiently strong dephasing of the coherences, the off-diagonal contributions in conjunction with the correction from the correlation contribution can be neglected in the equation of motion. Then, the carrier occupation adapts instantaneously to the new band structure. We assume that the composition of the bands does not change too fast and approximate the result of this adaptation to be $\tilde{n}^{b}_{\bvec{k}} \approx n^{b}_{\bvec{k}}$. We thus implement the time-dependent basis in the description of the carrier dynamics using time-dependent band-energies and wave-functions including time-dependent Coulomb-matrix elements due to the basis transformation with a time-dependent screening. The band dynamics here lead to an additional re-distribution of carriers into the new equilibrium distribution and a different pronunciation of intra- and inter-band scattering pathways.

For example the Coulomb scattering terms for the occupation $\tilde{n}^{b}_{\bvec{k}}$ at time $t_1$ can be written as
\begin{equation}
\frac{d}{dt} \tilde{n}^{b}_{\bvec{k}} = \frac{2\pi}{\hbar}
\sum_{\bvec{k}_{2}\bvec{k}_{3}}
\sum_{b_{2}b_{3}b_{4}}  \widetilde{\hat{W}} \big[ \tilde{N}^{\text{in}} - \tilde{N}^{\text{out}} \big] \delta ( \Delta\tilde{\epsilon}) 
\label{app_eom8}
\end{equation}
with 
\begin{align}
\widetilde{\hat{W}} &=& \tilde{W}^{b_{}b_{2}b_{3}b_{4}}_{\bvec{k}_{}\bvec{k}_{2}\bvec{k}_{3}\bvec{k}_{4} } \left( \tilde{W}^{b_{}b_{2}b_{3}b_{4}\text{*}}_{\bvec{k}_{}\bvec{k}_{2}\bvec{k}_{3}\bvec{k}_{4} } -
\tilde{W}^{b_{}b_{2}b_{4}b_{3}\text{*}}_{\bvec{k}_{}\bvec{k}_{2}\bvec{k}_{4}\bvec{k}_{3}} \right) \\
\tilde{N}^{\text{in}} &=& \left( 1 - \tilde{n}^{b}_{\bvec{k}} \right) \tilde{n}^{b_{2}}_{\bvec{k}_{2}}
\left( 1 - \tilde{n}^{b_{3}}_{\bvec{k}_{3}} \right) \tilde{n}^{b_{4}}_{\bvec{k}_{4}} \\
\tilde{N}^{\text{out}} &=& \tilde{n}^{b}_{\bvec{k}} \left( 1 - \tilde{n}^{b_{2}}_{\bvec{k}_{2}} \right)
\tilde{n}^{b_{3}}_{\bvec{k}_{3}} \left( 1 - \tilde{n}^{b_{4}}_{\bvec{k}_{4}} \right) \\
\Delta\tilde{\epsilon} &=& \tilde{\epsilon}^{b}_{\bvec{k}} -
\tilde{\epsilon}^{b_{2}}_{\bvec{k}_{2}} + \tilde{\epsilon}^{b_{3}}_{\bvec{k}_{3}} - \tilde{\epsilon}^{b_{4}}_{\bvec{k}_{4}} 
\end{align}

\bibliography{mybib}{}

\end{document}